\documentclass[prd,aps,nofootinbib,twocolumn,floatfix,10 pt]{revtex4}
\usepackage{graphicx}  
\usepackage{dcolumn}   
\usepackage{bm}        
\usepackage{amssymb}   
\usepackage{amsmath}
\usepackage{color,epsfig}
\usepackage{slashed}
\usepackage{epstopdf}
\usepackage{array}
\usepackage{dcolumn}
\usepackage[toc,page]{appendix}

\pdfoutput=1

\newcolumntype{P}[1]{>{\centering\arraybackslash}p{#1}}

\newcommand{\beqa}{\begin{eqnarray}}
\newcommand{\eeqa}{\end{eqnarray}}
\def\nn{\nonumber}

\def\roughly#1{\mathrel{\raise.3ex\hbox
{$#1$\kern-.75em\lower1ex\hbox{$\sim$}}}}

\def\gsim{\roughly>}
\def\bea{\begin{eqnarray}}
\def\eea{\end{eqnarray}}
\begin{document}

\title{ LHCb anomaly in $\boldsymbol{B\to K^*\mu^+ \mu^-}$ optimised
observables and potential of $\boldsymbol{Z^\prime}$ Model}
\author{Ishtiaq Ahmed$\footnote{ishtiaqmusab@gmail.com}^{1}$, Abdur Rehman$\footnote{abdur.rehman@fuw.edu.pl}^{1}$}
\affiliation{ $^{1}$National Centre for Physics, Quaid-i-Azam
University Campus,
Islamabad, 45320 Pakistan}

\date{\today}
\begin{abstract}
Over the last few years LHCb with present energies found some discrepancies in
$b\to s\ell^+\ell^-$ FCNC transitions including anomalies in
the angular observables of $B\to K^*\mu^+\mu^-$, particularly in $P_5^\prime$, in low dimuon mass
region. Recently, these anomalies are confirmed by Belle, CMS and ATLAS. As the direct evidence of physics
beyond-the-SM is absent so far, therefore, these anomalies are being interpreted as indirect hint of new physics.
In this context, we study the implication of  non universal family of $Z^\prime$
model to the angular observables $P_{1,2,3}$,  $P^{\prime}_{4,5,6}$ and
newly proposed lepton flavor universality violation observables, $Q_{4,5}$, in
 $B\to K^*(\to K\pi)\mu^+ \mu^-$ decay channel in the low dimuon mass region.
To see variation in the values of these observables from
their standard model values, we have chosen the different scenarios of the $Z^\prime$ model.
 It is found that these angular observables are sensitive to the values of the
parameters of $Z^\prime$ model. We have also found that with the present
parametric space of $Z^\prime$ model, the $P_5^\prime$-anomaly could be accommodated.
However, more statistics on the anomalies in the angular observables
are helpful to reveal the status of the considered model and, in general, the nature of new physics.
\end{abstract}
\maketitle



\section{Introduction}\label{intro}

In flavor physics, the study of rare $B$ meson decays provide us a powerful tool, not only
to test the standard model (SM) at loop level but also to search the possible
new physics (NP). searching of NP in rare decays of B-meson
demands to focus on those observables which contain minimum hadronic
 uncertainties such that they can be predicted precisely in the SM and are
available at current colliders. In exclusive rare B meson decays, the main source of hadronic uncertainties
come from the form factors which are non-perturbative quantities and are
difficult to compute. In addition, these uncertainties may preclude the signature of any possible NP.
From this point of view, among all rare decays, the four body decay channel,
$B\to K^*(\to K\pi)\mu^+\mu^-$, have a special interest in literature due
to the fact that it gives a large variety of angular observables, namely, $P_{i}$ $(i=1,2,3)$ and $P_i^\prime$ $(i=4,5,6)$ \cite{DescotesGenon:2012zf} which are
free from hadronic uncertainties \cite{Descotes-Genon:2013vna}. The comparison between the
theoretical predictions of these kind of observables in the SM with
the experimental data could be helpful to clear some smog on the physics beyond the SM.

From experimental point of view, few years back, LHCb measured the values of these angular observables for the decay channel $B\to K^*(\to K\pi)\mu^+\mu^-$. These measurements found a 3.7$\sigma$ deviation
in the value of $P_5^\prime$,  with 1 fb$^{-1}$ luminosity in the $s\in[4.30,8.68]$ GeV$^2$ bin \cite{2013lhcb}.
Recently, this discrepancy again seen at
LHCb with a 3$\sigma$ deviation with 3 fb$^{-1}$ luminosity in comparatively two shorter
 adjacent bins $s\in[4,6]$ GeV$^2$ \cite{2015lhcb} and  $s\in[6,8]$ GeV$^2$ which is also
confirmed by Belle in the larger bin   $s\in[4,8]$ GeV$^2$ \cite{belle2016, Wehle:2016yoi}.
The very recent results from ATLAS \cite{ATLAS-CONF-2017-023} and CMS \cite{CMS-PAS-BPH-15-008, Khachatryan:2015isa}  collaborations, presented in Moriond 2017, are also confirmed this discrepancy.
Furthermore, LHCb also found  2.6$\sigma$ deviation in the
value of $R_K=$ Br$(B\to K\mu^+\mu^-)/$Br$(B\to Ke^+e^-)$ \cite{rk}, and $\gsim2\sigma$ in
the Br$(B_s\to\phi\mu^+\mu^-)$ \cite{lhcbphi}. Interestingly, all these deviations
belong to the flavor changing neutral current
(FCNC) transitions, $b\to s\ell^+\ell^-$, where $\ell^-$ denotes the final state leptons.

The anomalies, mentioned above, are slowly piled up and received a considerable
attention in the literature (see for instance \cite{Matias:2012xw, Altmannshofer:2017fio}). It is also important to mention here that even the angular observables are form factor
independent (FFI) but for precise theoretical predictions,
one needs to incorporate the factorizable and non-factorizable QCD corrections.
The factorizable corrections absorb in the hadronic form factors
while the non-factorizable corrections arise from hard scattering of
the process and do not belong to the form factors.
 In this respect, there are some studies which focus to the question
whether these anomalies emerge from unknown factorizable
power corrections or from NP \cite{Capdevila:2017ert, Chobanova:2017ghn}. However,
global fit analysis with present data, strongly pointed out that the interpretation of mentioned
 anomalies through the NP is a valid option \cite{Altmannshofer:2017fio}. In the present study, to
determine the values of angular observables, we have included both type
of corrections up to next-to-leading order (NLO) and their expressions
are given in Appendix \ref{TransAmp}.



From NP point of view, several extensions of the SM have been put forwarded
\cite{models,Egede:2010zc,Altmannshofer14,bobeth13,thurth14,jager14,Allanach:2015gkd,Crivellin:2015era,Jager:2017gal}. Among these, the $Z^\prime$
model is economical due to the fact that besides the SM gauge group, it requires only one extra $U(1)^\prime$
gauge symmetry associated with a neutral gauge boson, called $Z^\prime$. The nature of couplings of the $Z^\prime$ boson with the
quarks and leptons leads the  FCNC transitions to the tree level. In this model, the NP effects comes only through the short distance
Wilson coefficients which are encapsulated in the new coefficients
$C_9^{\text{tot}}=C_9^{\text{SM}}+C_9^{Z^\prime}$, $C_{10}^{\text{tot}}=C_{10}^{\text{SM}}+C_{10}^{Z^\prime}$, while operator basis  remained unchange.

Several previous studies
shown a possible interpretation to alleviate the mismatch between the experimental data of different observables for the decay $B\to K^*\mu^+\mu^-$ and their SM predictions
in terms of $Z^\prime$ model \cite{Descotes-Genon:2013wba,
Altmannshofer:2013foa,Gauld:2013qba,Gauld:2013qja,Buras:2013dea,Buras:2013qja} without any conflict.
Therefore, it is natural to ask whether the $Z^\prime$ model could explain the recently
observed anomalies in the angular observables of the decay channel
$B\to K^*(\to K\pi)\mu^+\mu^-$. With this motivation, in the current study,
we have analyzed the optimal observables $P_{1,2,3}$ and $P^{\prime}_{4,5,6}$,
in the low dimuon mass region, for the $B\to K^*(\to K\pi)\mu^+\mu^-$ in
the SM and in the $Z^\prime$ model. Besides these observables, we have also calculated the
violation of lepton  flavor universality (LFU) observables
namely, $Q_{4(5)}=P^{\prime \mu}_{4(5)}-P^{\prime e}_{4(5)}$ \cite{Capdevila:2016ivx}. For numerical calculations
of these observables, we have used the LCSR values
of the hadronic form factors \cite{zwicky05} and for
$Z^\prime$ parameters, we have used the Utfit collaboration values, called as $\mathcal{S}_1$, $\mathcal{S}_2$ and another different scenario, called $\mathcal{S}_3$  which numerical values are listed in Tab. (\ref{ZP table}).

We would like to mention here that the considered scenarios labeled as $\mathcal{S}_1$, $\mathcal{S}_2$ and $\mathcal{S}_3$ have same coupling structure of the $Z^\prime$ boson with the quarks and the leptons. However, the underlying difference between these scenarios is related to the different fit values of parameters such as new weak phase and couplings of $Z^\prime$ model, for considered decay process, available in the literature. For example, by using the all available experimental data on $B_s-\bar B_s$ mixing, Utfit collaboration has found two solutions of new weak phase, $\phi_{sb}$, that arises due to the measurement ambiguities in the data and referred as $\mathcal{S}_1$ and $\mathcal{S}_2$. Similarly, another possible constraint on parameters of $Z^\prime$ model is discussed in \cite{newcon4} that, hereafter, we label as $\mathcal{S}_3$.

This paper is organized as follows: Section \ref{MatrixFF}, contains the
effective Hamiltonian for the $b\to s\ell^+\ell^-$ transition in the SM and in
the $Z^\prime$ model. The $B\to K^*$ matrix elements in terms of form factors
and the expression of differential decay distributions are also given
in this section. Formulae for the angular observables in section \ref{AnalyticAngObs}.
In section \ref{results}, we have plotted
the angular observables and their average values against dimuon mass $s$ and we
have given phenomenological analysis of these observables. In the last section
we conclude our work. Appendix A contains the analytical expressions of the angular observables and the values of
input parameters. The contributions of factorizable and non-factorizable
corrections at NLO are summarized in Appendix \ref{TransAmp}.

\section{Formulation for the Analysis}
\subsection{Matrix Elements and Form Factors}\label{MatrixFF}
In the standard model, FCNC transition, $b\to s\ell^+\ell^-$, occurs at loop level which amplitude can be written as,
 \begin{align}
&\mathcal{M}^{\text{SM}}(b \rightarrow s \ell^{+}\ell^{-})= -\frac{\alpha \, G_{F}}{2\sqrt{2}%
\pi } V_{tb}V_{ts}^{\ast }\notag \\
&\times\bigg\{\langle K^*(p_{K^*},\epsilon)|\bar{s}\gamma^\mu L b|B(p_B) \rangle( C_{9}^{\rm eff}\bar{\ell} \gamma ^{\mu
}\ell+C_{10}^{\rm SM}\bar{\ell}\gamma ^{\mu }\gamma
_{5}\ell) \notag\\
&
-2m_{b}C_{7}^{\rm eff}\langle K^*(p_{K^*},\epsilon)|\bar{s}i\sigma _{\mu \nu }\frac{q^{\nu }}{q^{2}}R b|B(p_B) \rangle\,\bar{ \ell%
}\gamma ^{\mu }\ell \bigg\},\label{quark-amplitude}
\end{align}
where $L,R=(1\mp\gamma^5)$, $p_{K^*}$ and $\epsilon$ are  momentum and
 polarization of $K^*$ meson, respectively, while $p_B$ is the momentum of $B$ meson.

In the presence of $Z^\prime$ the FCNC transitions could occur at tree
level and the Hamiltion can be written in the following form (see detail in the refs. \cite{PL1,QCh,KCh1,VB3} )
\begin{eqnarray}
\mathcal{H}^{Z^{\prime}}_{\rm eff}&=&-\frac{4G_{F}}{\sqrt{2}} V_{tb}V^{\ast}_{ts}\left[\Lambda_{sb}\,
C_{9}^{Z^{\prime}}O_{9}+\Lambda_{sb}\, C_{10}^{Z^{\prime}}O_{10}\right],\notag \\ \label{lamb1}\\
\text{where,\,\,\,} \Lambda_{sb}&=&\frac{4\pi e^{-i\phi_{sb}}}{\alpha_{em}V_{tb}V^{\ast}_{ts}}\label{lamb2},\,
 C_{9}^{Z^{\prime}}=|\mathcal{B}_{sb}|S^{LR}_{\ell\ell},  \notag\\
\text{and \quad} C_{10}^{Z^{\prime}}&=&|\mathcal{B}_{sb}|D^{LR}_{\ell\ell} \quad \text{with,\,\,\,} \notag \\
\vspace{10.6cm}
\mathcal{S}^{LR}_{\ell\ell}&=&\mathcal{B}^{L}_{\ell\ell}+\mathcal{B}^{R}_{\ell\ell},\quad
\mathcal{D}^{LR}_{\ell\ell}=\mathcal{B}^{L}_{\ell\ell}-\mathcal{B}^{R}_{\ell\ell}.\label{4}
\end{eqnarray}
The $\mathcal{B}_{sb}$ is the coupling of  $Z^\prime$ with  quarks and
$\mathcal{B}^{L}_{\ell\ell}$, $\mathcal{B}^{R}_{\ell\ell}$ are left and
right-handed couplings fo $Z^\prime$ with  leptons. One can notice from
 Eq. (\ref{lamb2}) that in the $Z^\prime$ model, operator basis remains
the same as in the SM while Wilson coefficients, $C_{9}$ and $C_{10}$, get modified.
The total amplitude for the decay $B\to K^*\ell^{+}\ell^{-}$  is the sum of SM and $Z^{\prime}$ contributions,
and can be written as follows,
\begin{eqnarray}
&\mathcal{M}^{tot}(B\to K^*\ell^{+}\ell^{-})
= -\frac{\alpha\, G_{F}}{2\sqrt{2}%
\pi } V_{tb}V_{ts}^{\ast } \notag \\
&\times\bigg\{\langle K^*(p_{K^*},\epsilon)|\bar{s}\gamma^\mu L b|B(p_B) \rangle( C_{9}^{\text{tot}}\bar{\ell} \gamma ^{\mu
}\ell+C_{10}^{\text{tot}}\bar{\ell}\gamma ^{\mu }\gamma
_{5}\ell) \notag\\
&-2m_{b}C_{7}^{\rm eff}\langle K^*(p_{K^*},\epsilon)|\bar{s}i\sigma _{\mu \nu }\frac{q^{\nu }}{q^{2}}R b|B(p_B) \rangle\bar\ell
\gamma^{\mu}\ell \bigg\},\label{Atot}
\end{eqnarray}
where $C_{9}^{\rm tot}=C_{9}^{\rm eff}+\Lambda_{sb} C_{9}^{Z^{\prime}}$ and $C_{10}^{\rm tot}=C_{10}^{\rm SM}+\Lambda_{sb} C_{10}^{Z^{\prime}}$.

The matrix elements for
$B\to K^*$ transition, appears in Eq.~(\ref{Atot}), can be written
in terms of form factors as follows
\bea
&&\langle K^*(p_{K^*},\epsilon)|\bar{s}\gamma^\mu L b|B(p_B) \rangle =  - iq_{\mu}
\frac{2m_{K^*}}{s} \epsilon^{*} \cdot q \notag \\
&&\times \bigg[
A_3(s)-A_0(s) \bigg]-\epsilon_{\mu\nu\lambda\sigma} \epsilon^{*\nu}
p^{\lambda}_{K^*} q^{\sigma} \frac{2V(s)}{(m_B+m_{K^*})} \nn \\
&&  + i\epsilon_{\mu}^{*}(m_B+m_{K^*})
A_1(s) \notag \\
&&\mp i(p_B+p_{K^*})_{\mu}\epsilon^{*} \cdot q
 \frac{A_2(s)}{(m_B+m_{K^*})}, \notag
\eea
\bea
&&\langle K^*(p_{K^*},\epsilon)|\bar{s}i\sigma_{\mu\nu}q^\nu R b|B(p_B) \rangle  =  2
\epsilon_{\mu\nu\lambda\sigma} \epsilon^{*\nu}
p^\lambda_{K^*} q^\sigma ~ T_1 (s)\notag \\
&&+~i \epsilon^{*} \cdot q
 \bigg\{ q_\mu -
\frac{(p_B + p_{K^*})_\mu s}{(m_B^2-m_{K^*}^2)} \bigg\}
T_3(s) \nn \\
&&  +~i \bigg\{
\epsilon^*_{\mu}(m_B^2-m_{K^*}^2)-(p_B+p_{K^*})_\mu
 \epsilon^{*} \cdot q
 \bigg\}~T_2(s),
\eea
where,
\bea
A_3(s)&=&\frac{m_B+m_{K^*}}{2m_{K^*}}A_1(s)-\frac{m_B-m_{K^*}}{2m_{K^*}}A_2(s)\;.
\eea
Here $A_{0,1,2}(s)$, $V(s)$, $T_{1,2,3}(s)$ are the form factors and contain hadronic  uncertainties.
At leading order by using  the heavy quark limit,
the QCD form factors follow the symmetry relations and can be expressed in terms of
two universal form factors $\xi_\perp$ and $\xi_\parallel$ \cite{hiller2,Beneke:2000wa}.
\begin{align}
  \label{eq:xi:def}
  \xi_\perp & = \frac{m_B}{m_B + m_{K^*}} V, \notag \\
    \xi_\parallel & = \frac{m_B + m_{K^*}}{2 E_{K^*}} A_1 -
                    \frac{m_B - m_{K^*}}{m_B} A_2.
\end{align}
It is also important to mention here that the angular observables are soft form factor independent at LO in $\alpha_s$ (i.e., not totally dependent of FF). There is residual dependence has been discussed, computed systematically and included in the predictions of the main papers of the field and even if, as expeted, does not play an important role induce certain mild dependence on FF. In addition, for the $s$ dependence of the universal form factors there are different parametrization \cite{Straub:2015ica}, however, we have analyzed that the choice of parametrization is not so important at low $s$. In the current study, we use the following parametrization of
 LCSR approach \cite{zwicky05}.
\begin{align}
  V(s) & = \frac{r_1}{1 - s/m_R^2} + \frac{r_2}{1 - s/m_{fit}^2} ,
  A_1(s)  = \frac{r_2}{1 - s/m_{fit}^2} ,\notag
\\
  A_2(s) & = \frac{r_1}{1 - s/m_{fit}^2} + \frac{r_2}{(1 - s/m_{fit}^2)^2} ,\label{form}
\end{align}
where the  parameters $r_{1,2}$, $ m^2_{R}$ and $m^2_{fit}$  are listed in Tab. (\ref{TabForm}).
The uncertainty in the universal form factors $\xi_\perp$ and $\xi_\parallel$ arises
from the uncertainty in the different parameters using in LCSR approach which is
about  $11\%$ and $14\%$, respectively, as discussed in \cite{hiller2}.
\begin{table}[hb]
\caption{\sf The values of the fit parameters involved in the calculations of the
form factors given in  Eq. (\ref{form}) \cite{zwicky05}. }\label{TabForm}
\centering
\begin{tabular}{|c|cccc|}
\hline
   & $r_1$ & $r_2$ & $m_R^2(\text{GeV}^2)$ & $m_{fit}^2 (\text{GeV}^2)$   \\
\hline\hline
   $ V(s)$&0.923&-0.511&28.30&49.40\\
$A_1(s)$&&0.290&&40.38\\
$A_2(s)$&-0.084&0.342&&52.00\\
\hline
\end{tabular}
\end{table}
At NLO, the relations between the $T_i(s)$ where ($i=1,2,3$) and the
invariant amplitudes $\mathcal{T}_{\perp,\parallel}(s)$, where $\mathcal{T}_{\perp,\parallel}=\mathcal{T}^-_{\perp,\parallel}$,  read as \cite{Beneke:2001at}.
\begin{eqnarray}
T_1(s)&=&\mathcal{T}_\perp,\quad T_2(s)=\frac{2\mathcal{E}_{K^*}}{m_B}\mathcal{T}_\perp, \quad
T_3(s)=\mathcal{T}_\perp+\mathcal{T}_\parallel,\notag \\
\end{eqnarray}
where $\mathcal{E}_{K^*}=(m_B^2+m_{K^*}^2-s)/2m_B$ is the energy of kaon in the rest frame of  $B$-meson and  $\mathcal{T}_{\perp,\parallel}(s)$ are defined in   Eq. (\ref{B5}) of Appendix \ref{TransAmp}.
\begin{figure*}[ht]
\begin{center}
\begin{tabular}{lcr}
\includegraphics[width=80mm,angle=0]{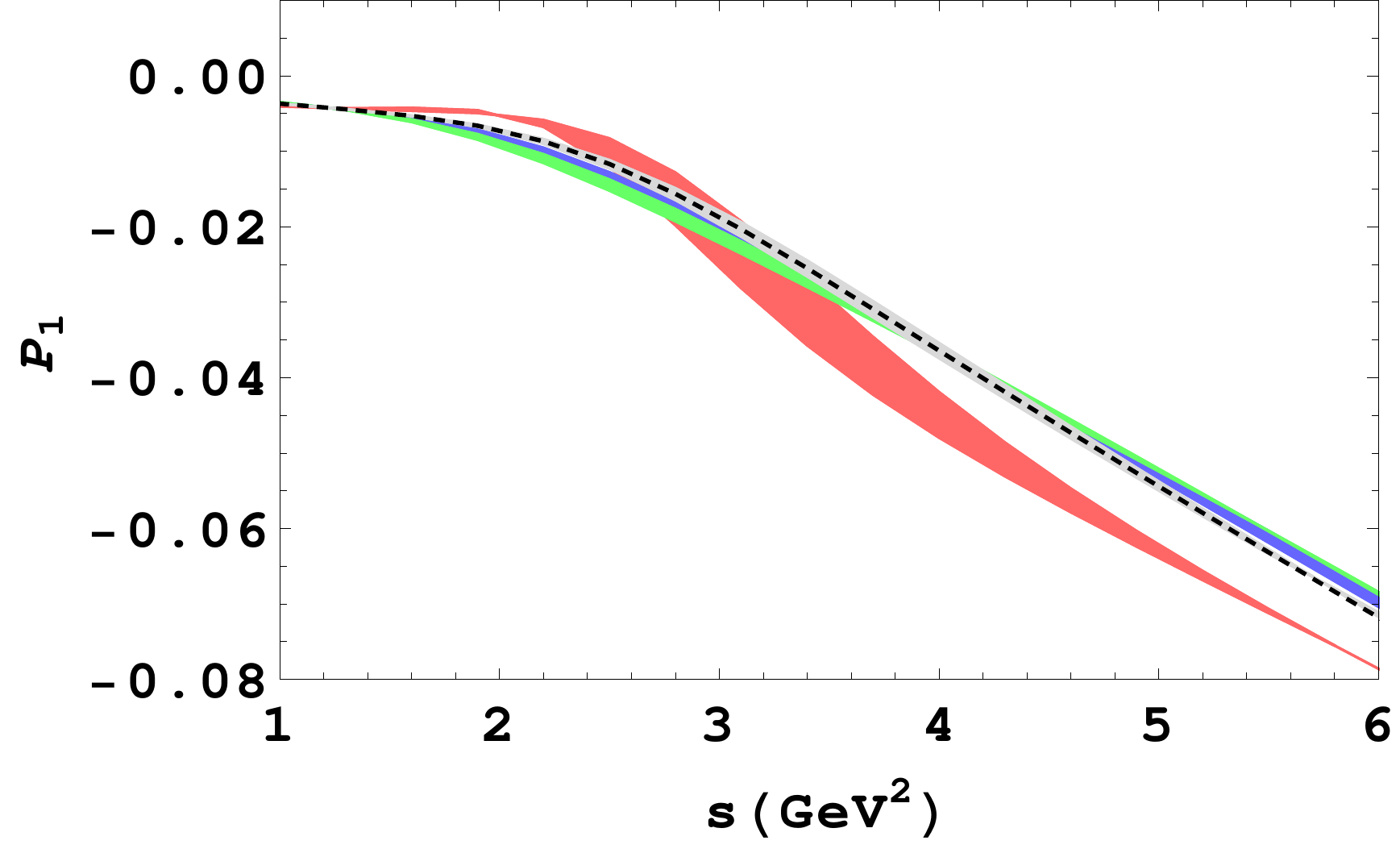}
\includegraphics[width=78mm,angle=0]{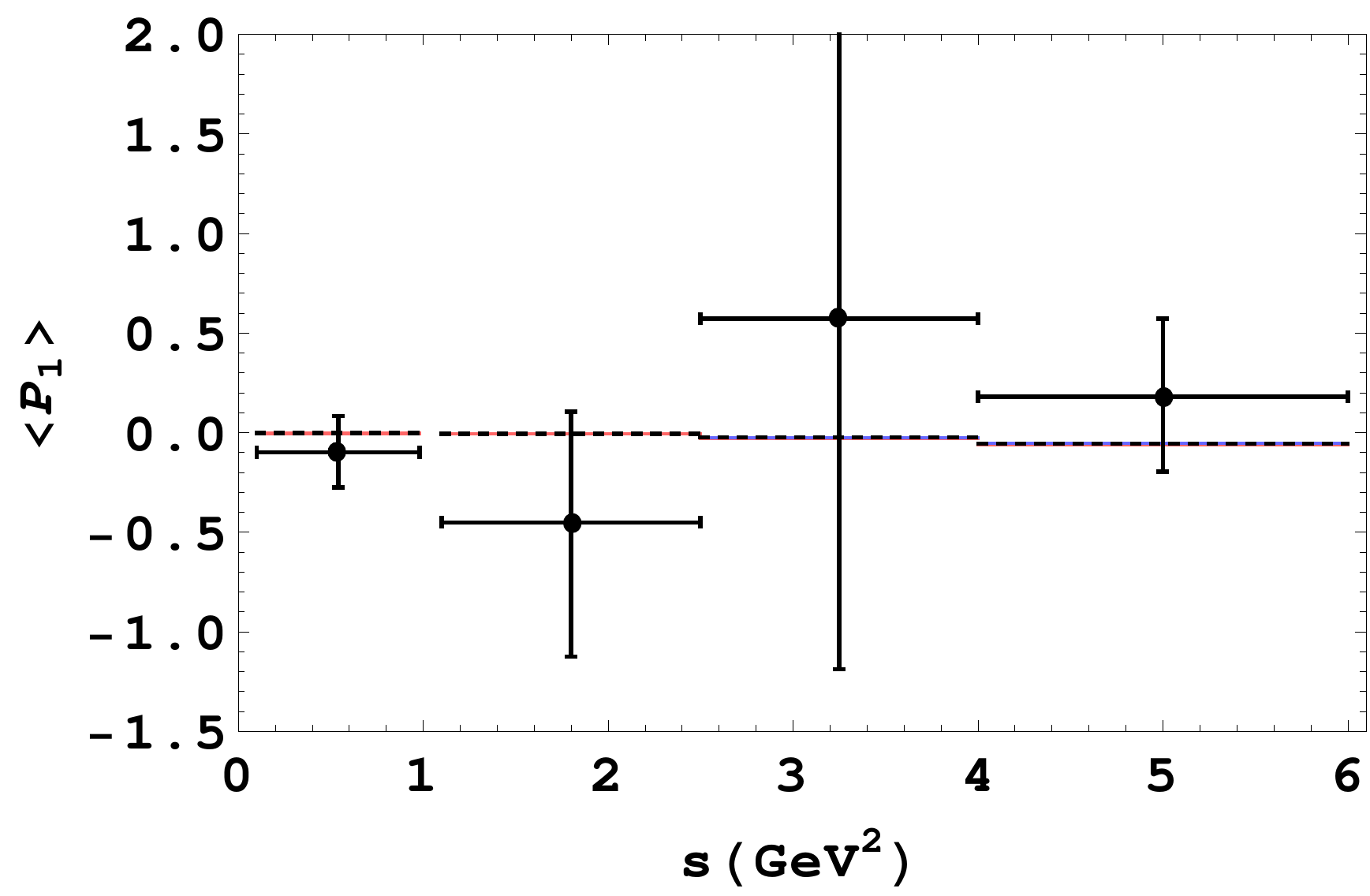}\\
\includegraphics[width=82mm,angle=0]{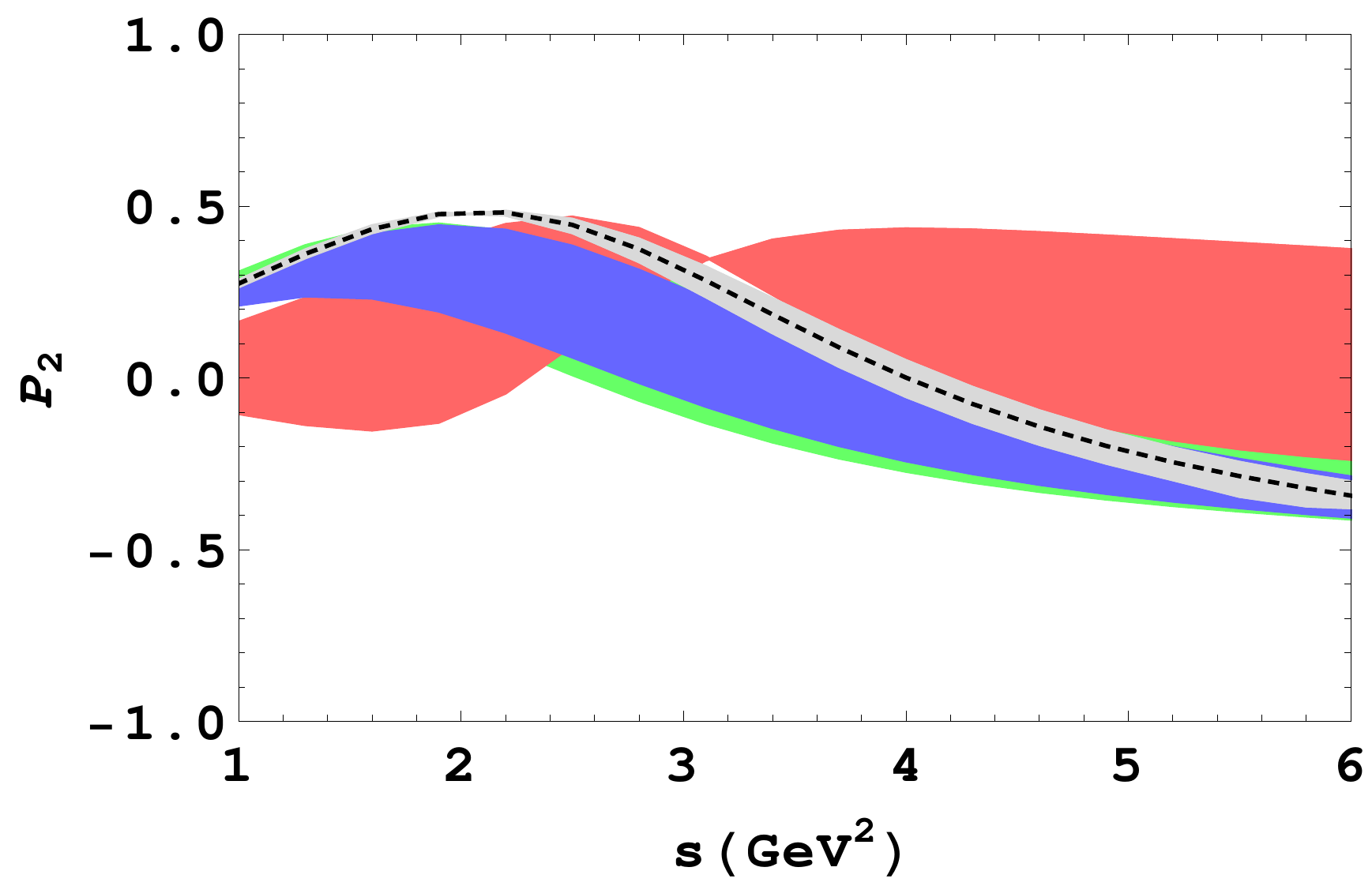}
\includegraphics[width=78mm,angle=0]{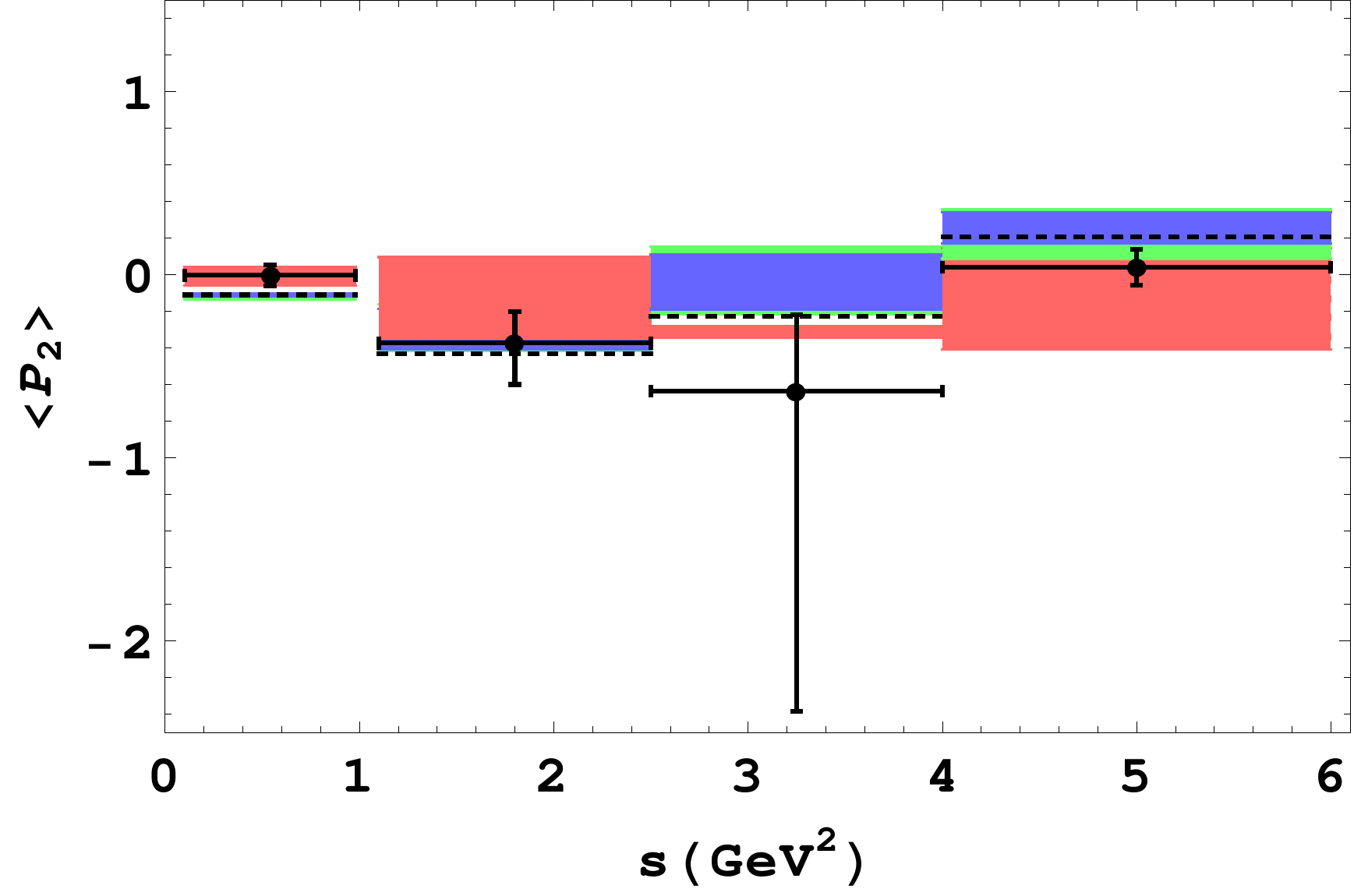}\\
\includegraphics[width=80mm,angle=0]{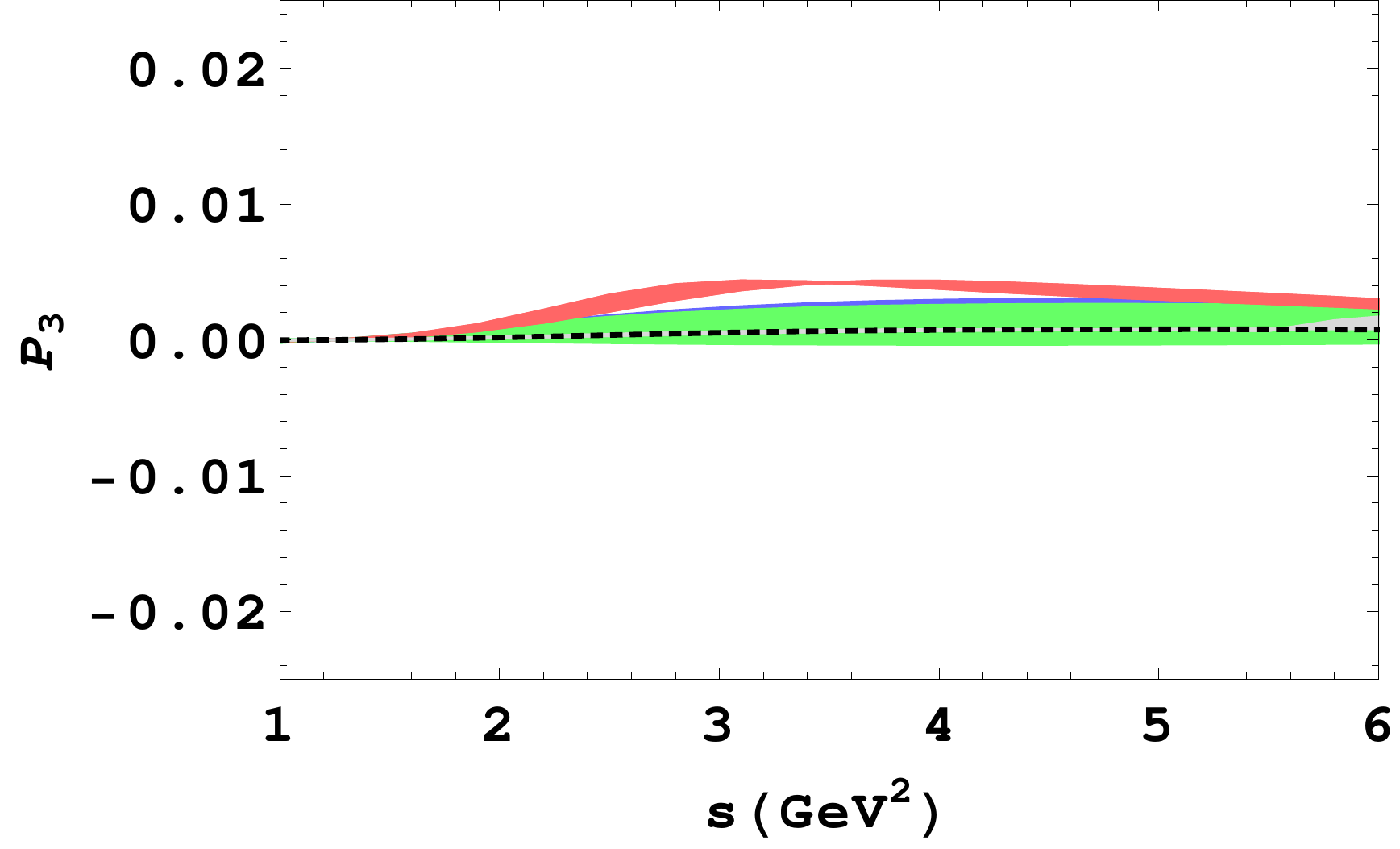}
\includegraphics[width=76mm,angle=0]{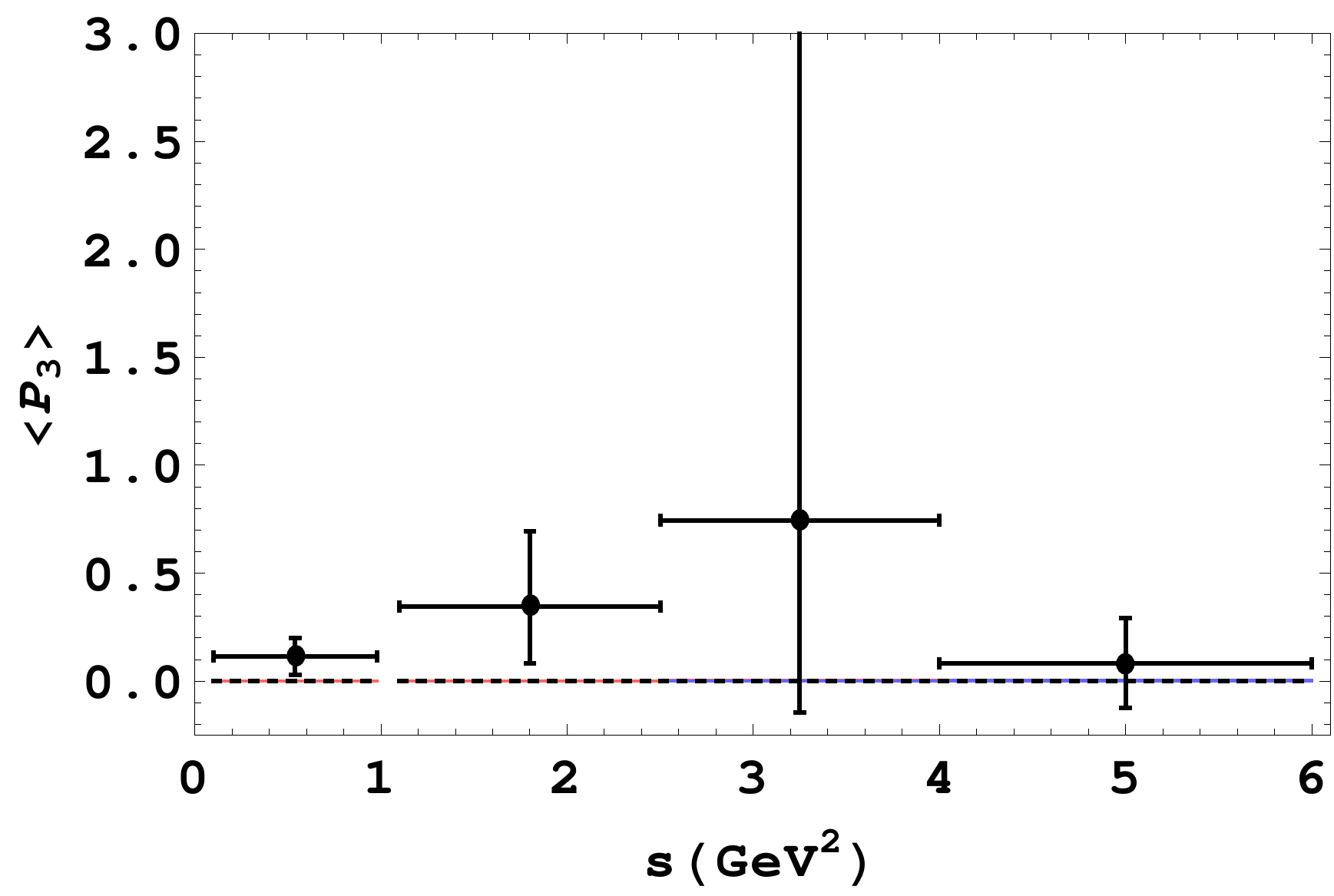}
\end{tabular}
\caption{\sf The dependence of the optimal observables, $P_{1,2,3}$ and  $\langle P_{1,2,3}\rangle$  for
the decay $B\to K^*(\to K\pi)l^+l^-$ on $s$. The black dashed line correspond to the SM while green,
blue and red bands correspond to the $\mathcal{S}_1$, $\mathcal{S}_2$ and $\mathcal{S}_3$
scenarios of the $Z^\prime$ model, respectively.} \label{p1p}
\end{center}
\end{figure*}
\begin{figure*}[ht]
\begin{tabular}{lcr}
\includegraphics[width=80mm,angle=0]{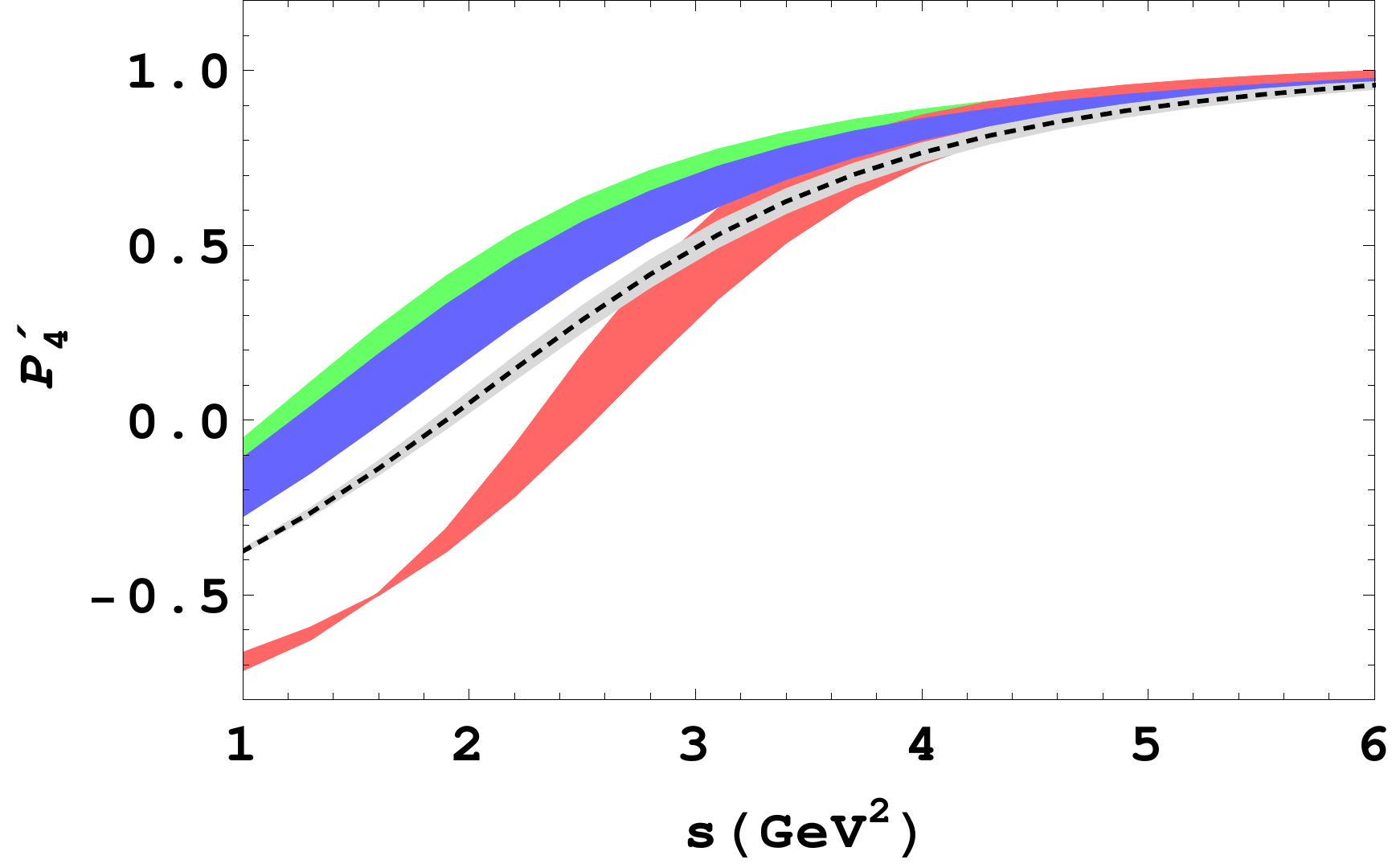}
\includegraphics[width=80mm,angle=0]{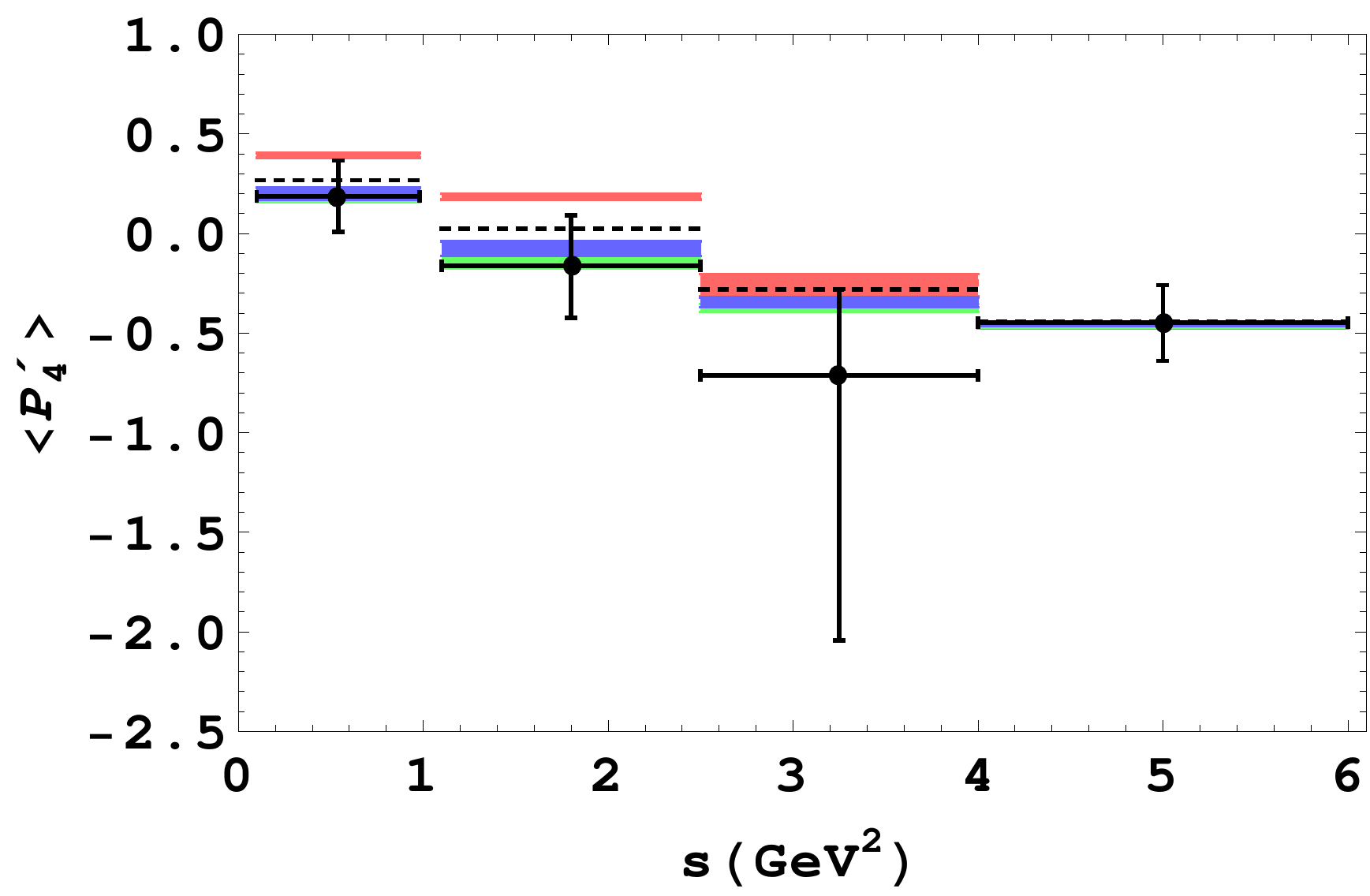}\\
\includegraphics[width=80.5mm,angle=0]{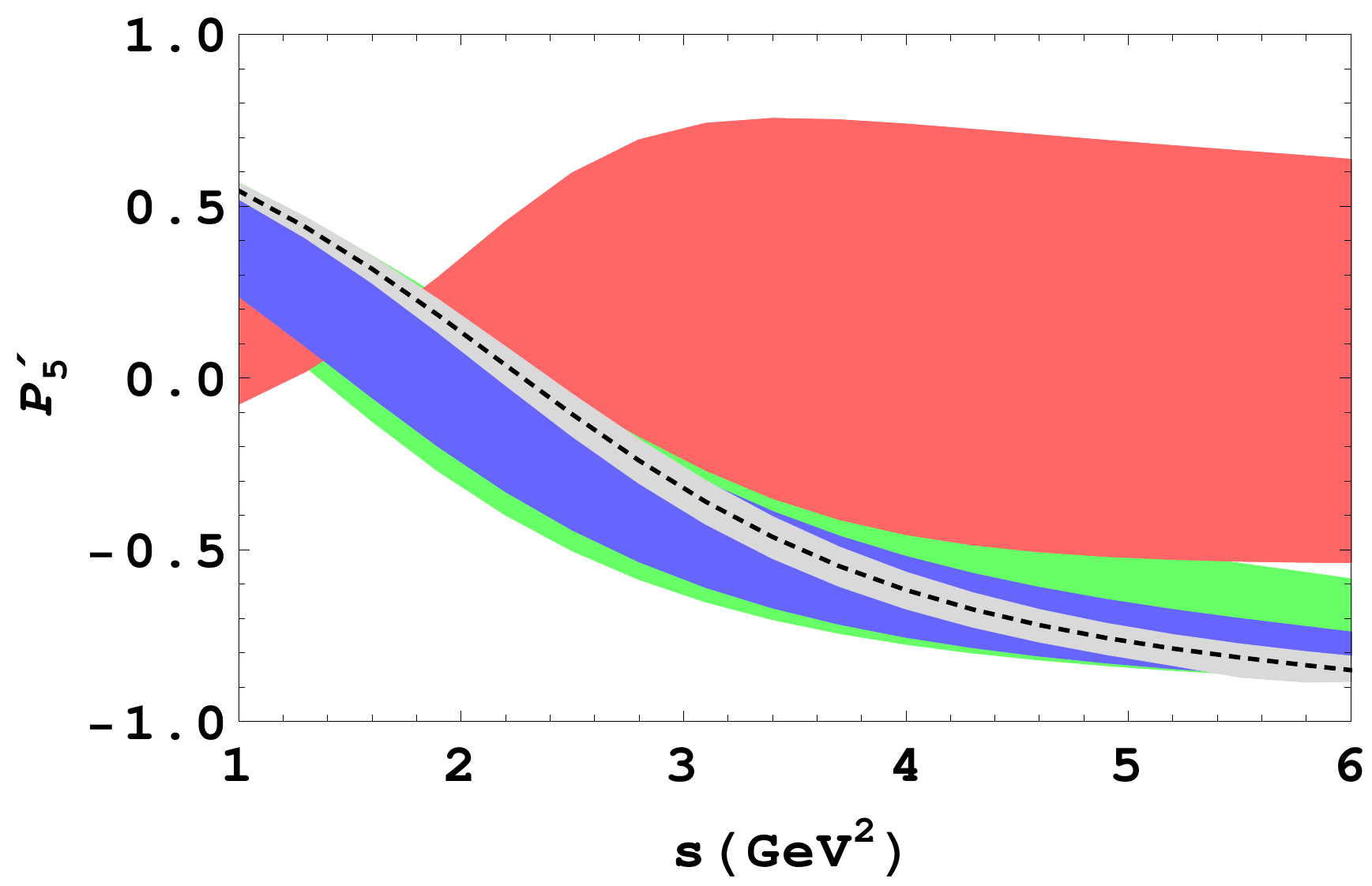}
\includegraphics[width=81mm,angle=0]{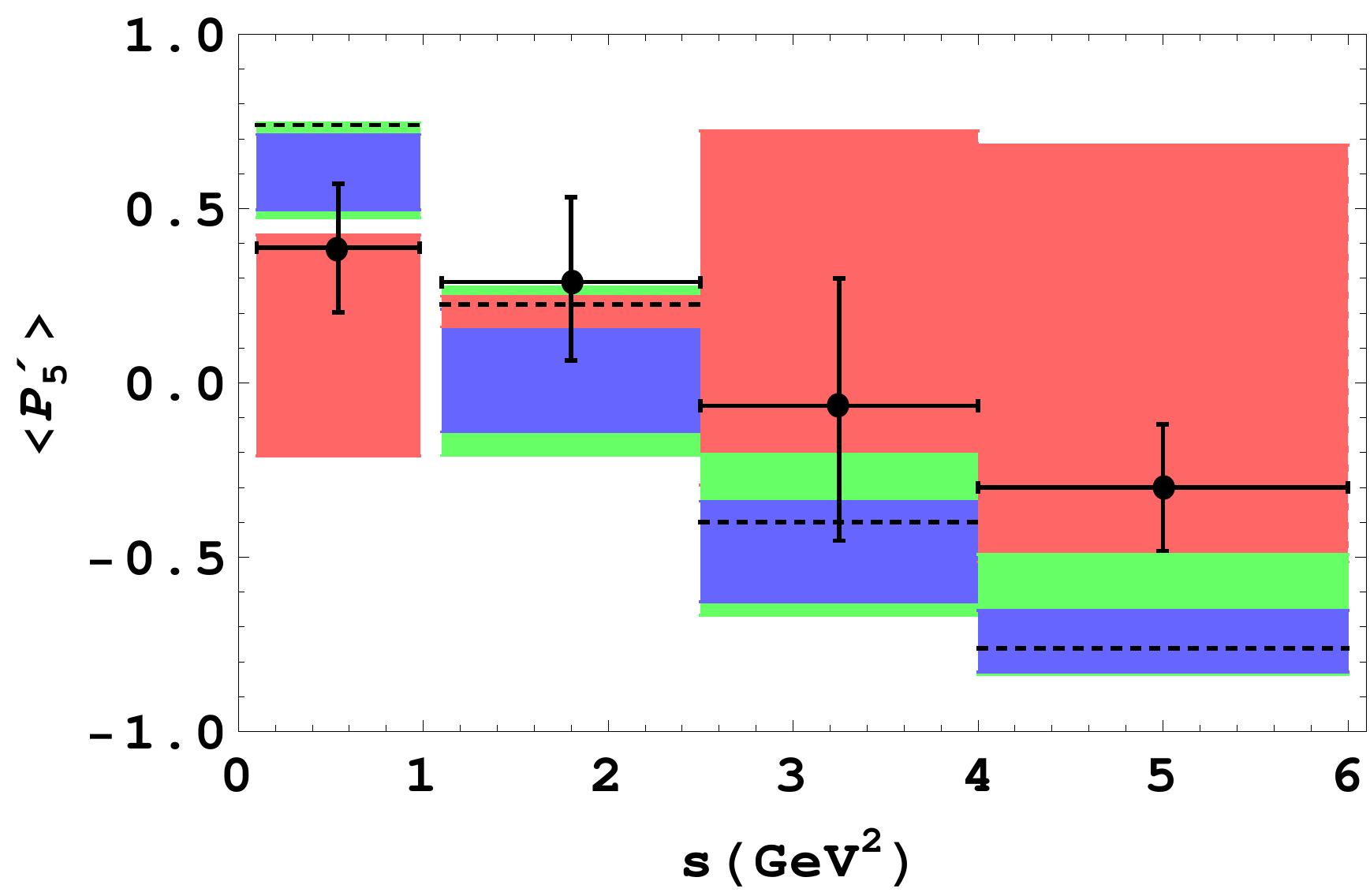}\\
\includegraphics[width=80mm,angle=0]{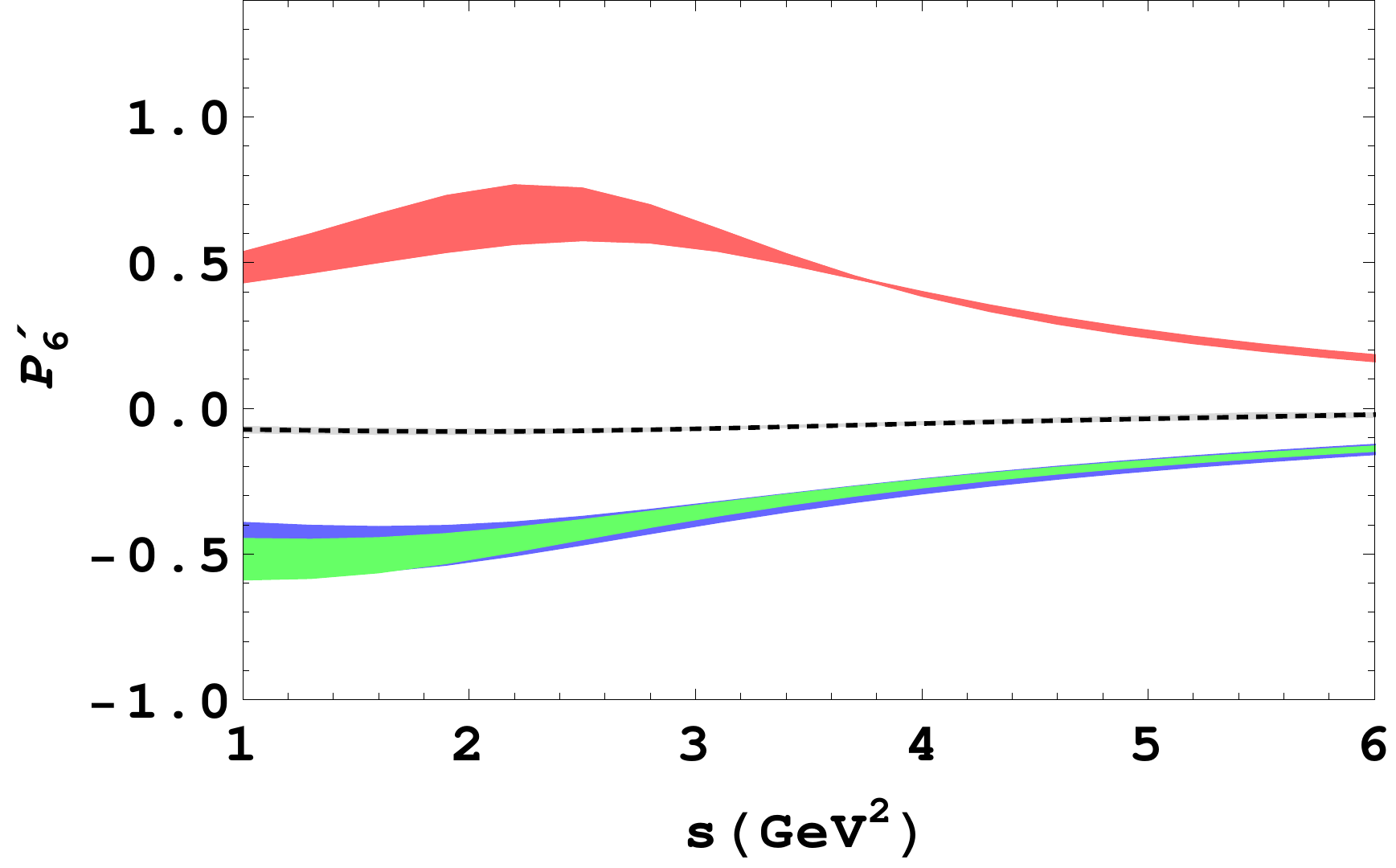}
\includegraphics[width=80mm,angle=0]{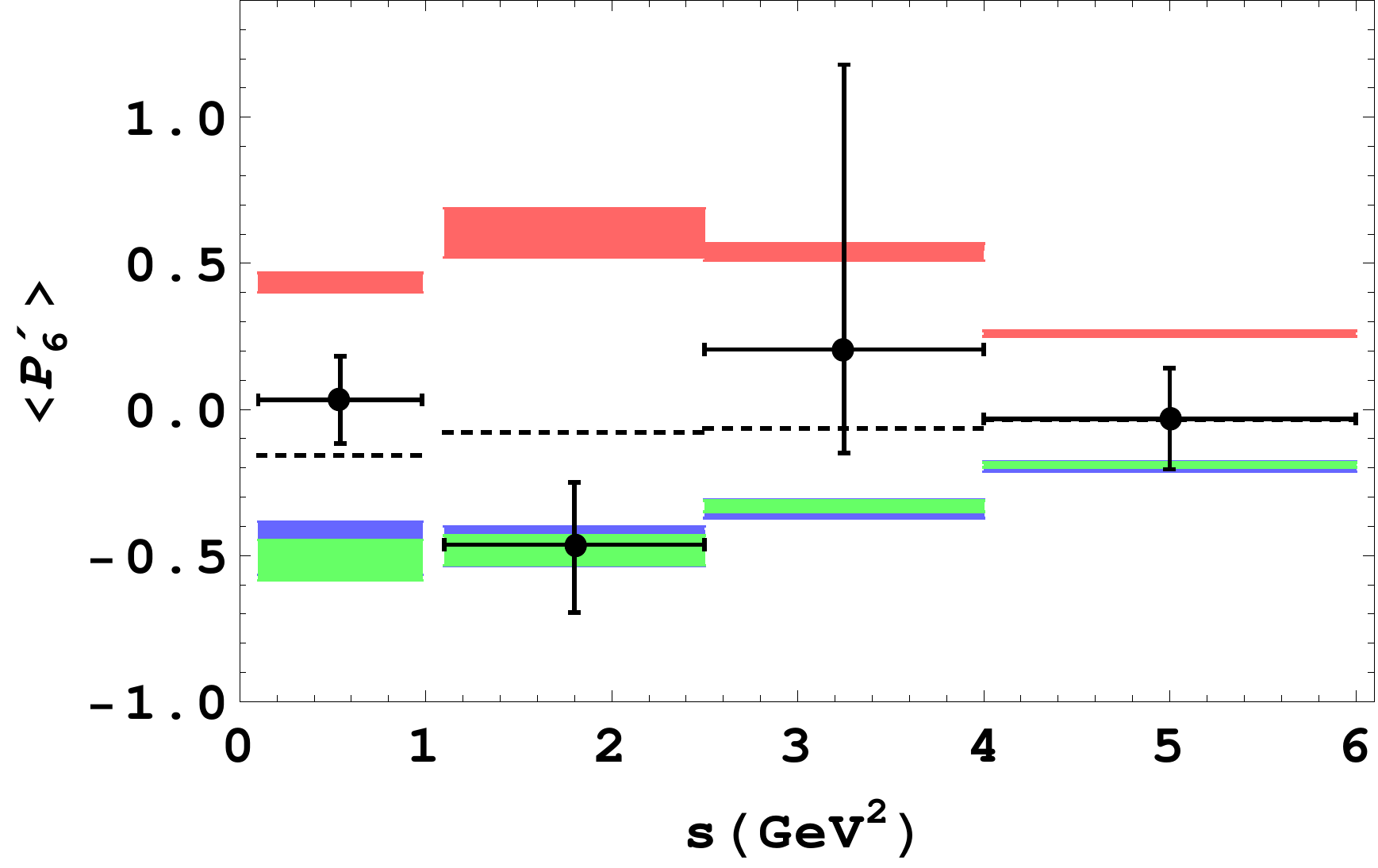}
\end{tabular}
\caption{\sf The dependence of the optimal observables, $P^\prime_{4,5,6}$ and  $\langle P^\prime_{4,5,6}\rangle$ for
the decay $B\to K^*(\to K\pi)l^+l^-$ on $s$, the legends are same as in Fig. (\ref{p1p}).} \label{p2p}
\end{figure*}
\begin{figure}
\begin{tabular}{lcr}
\includegraphics[width=80mm,angle=0]{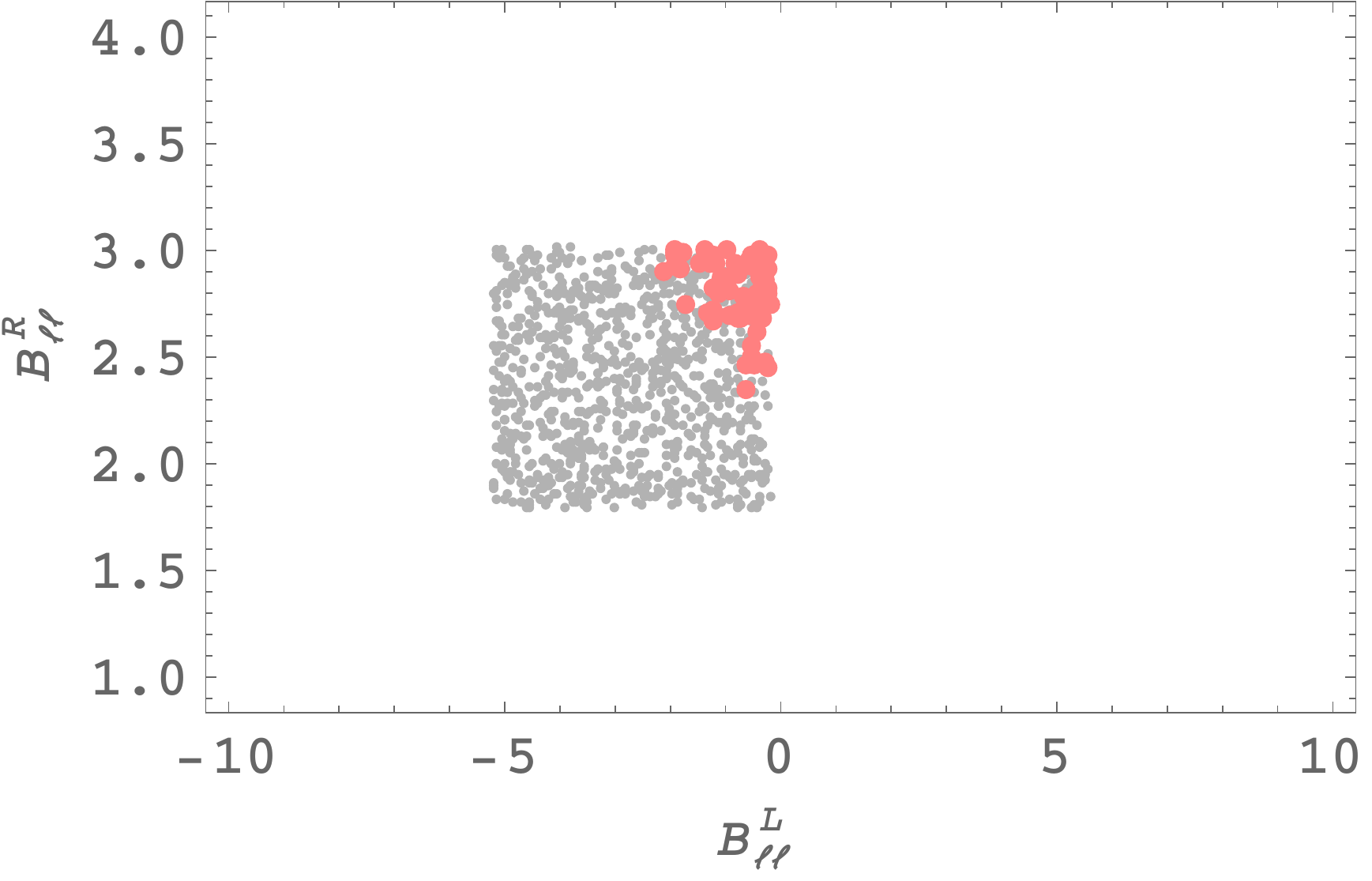}
\end{tabular}
\caption{Gray dots represent the left (right) couplings, ($B_{\ell\ell}^L$, $B_{\ell\ell}^R$), of $Z^\prime$ with leptons in $\mathcal{S}_1$ while the red dots show the values of these couplings after accommodating the $P_5^\prime$ anomaly in the $s\in[4.0, 6.0]$ GeV$^2$}  \label{constCou}
\end{figure}
\begin{figure*}[ht]
\begin{center}
\begin{tabular}{lcr}
\includegraphics[width=80mm,angle=0]{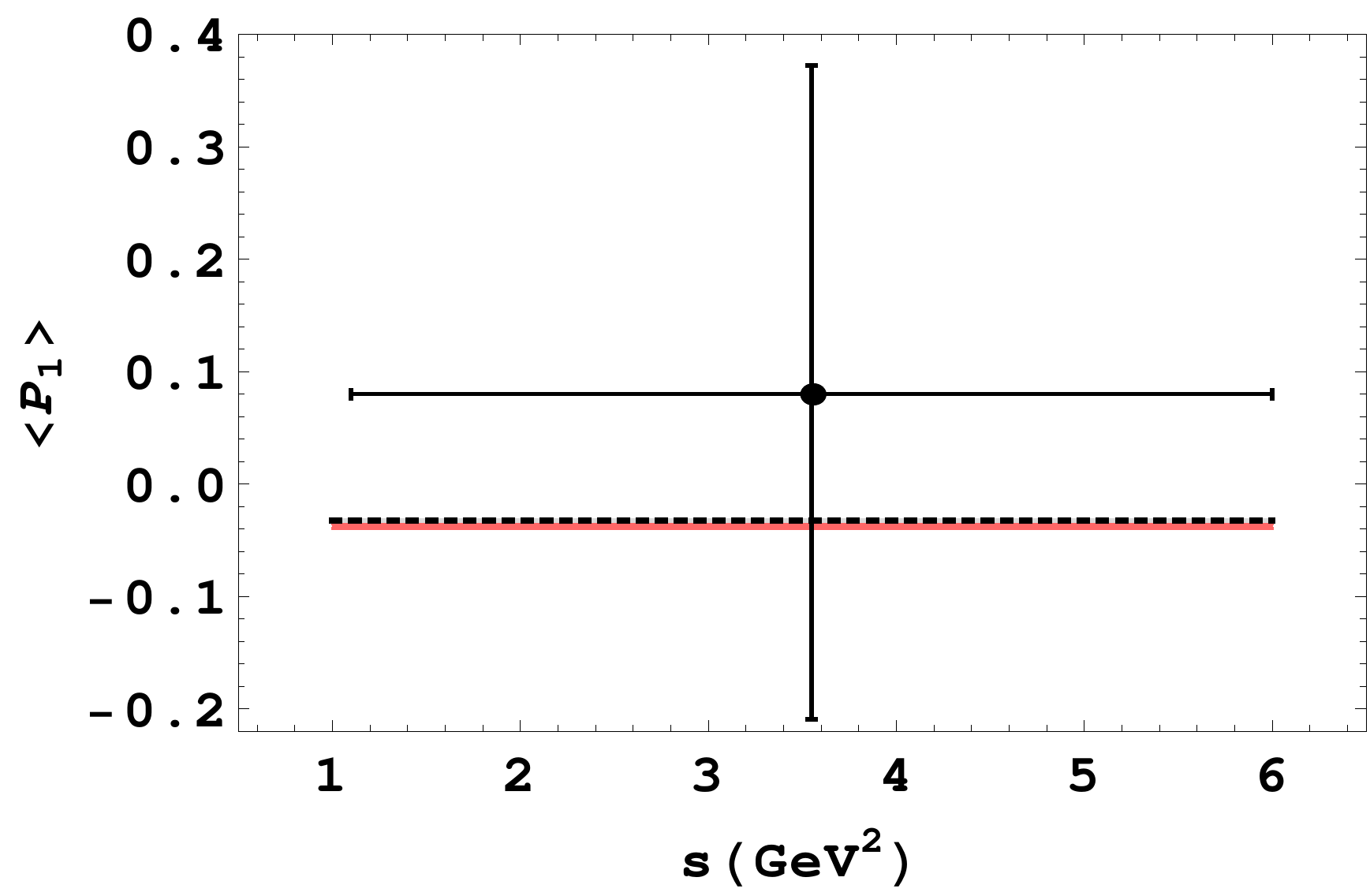}
\includegraphics[width=80mm,angle=0]{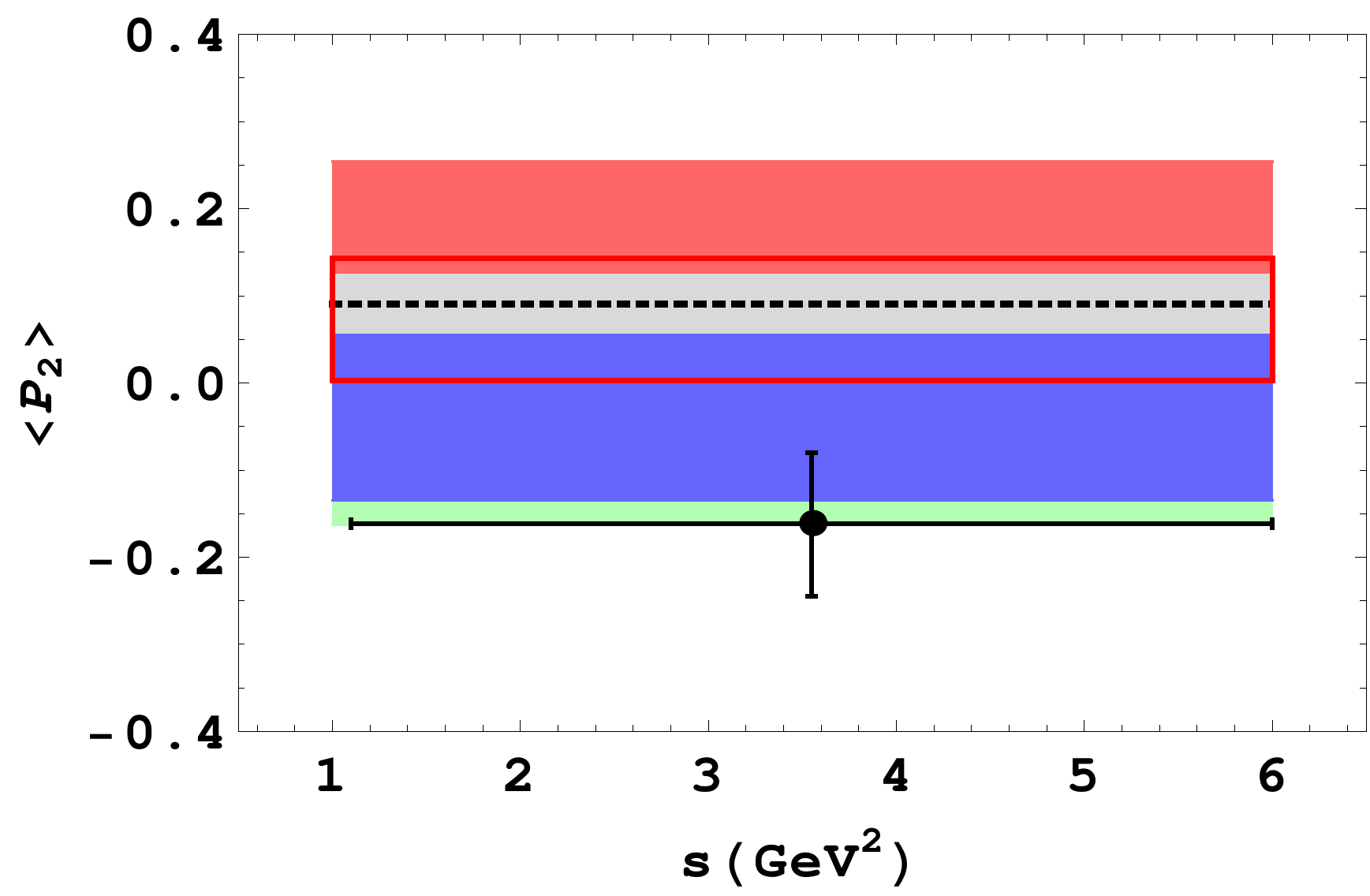}\\
\includegraphics[width=80.4mm,angle=0]{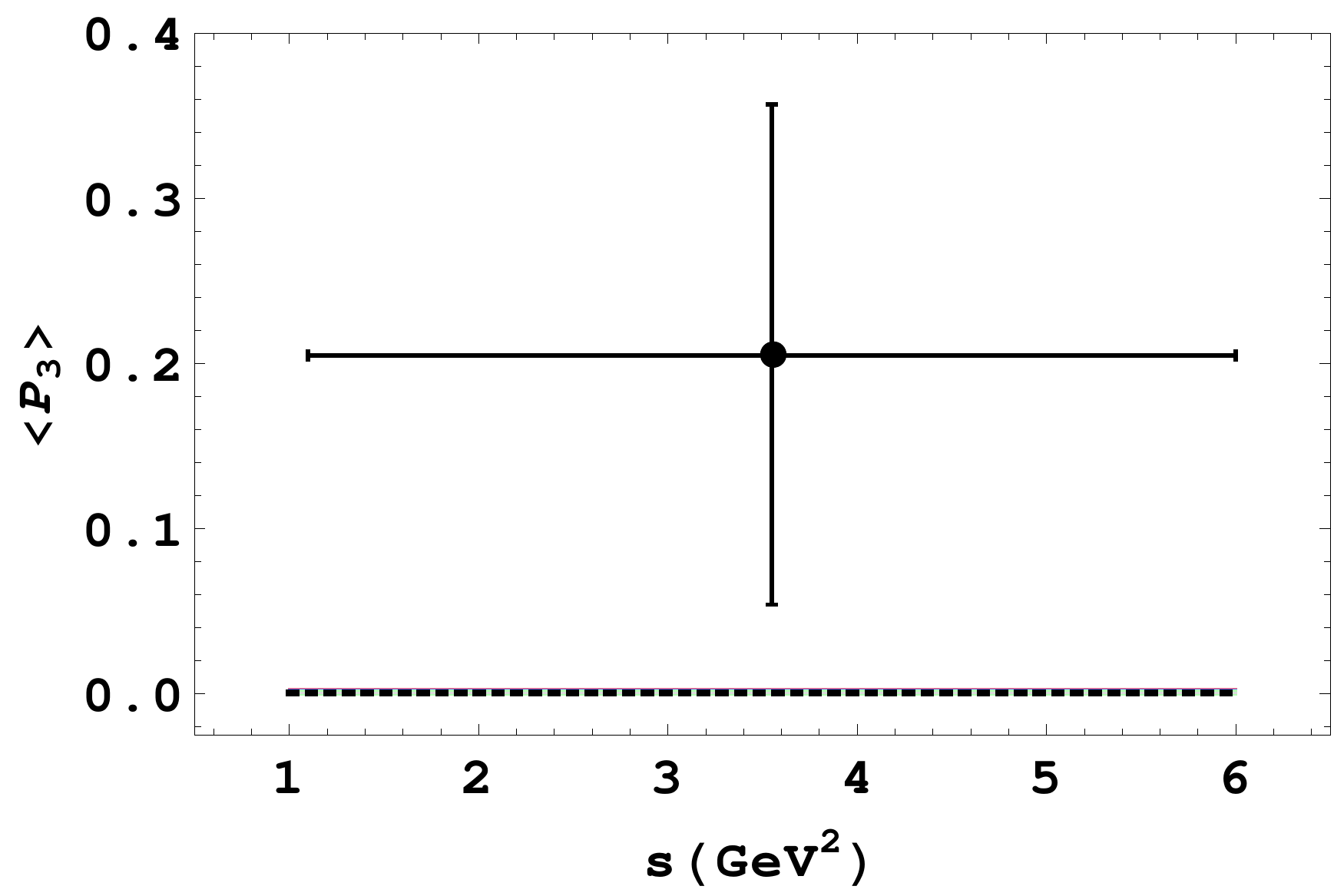}
\includegraphics[width=80mm,angle=0]{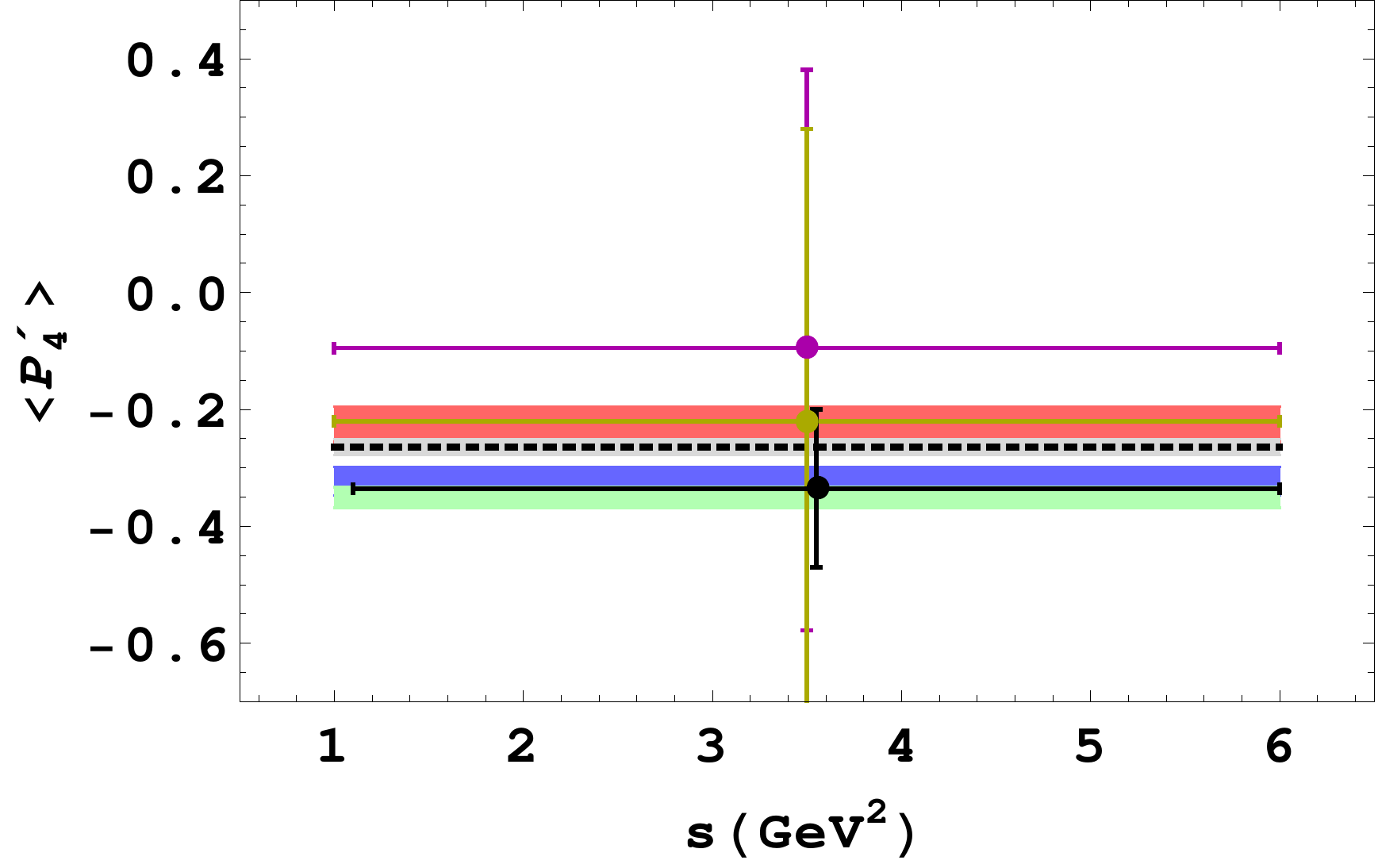}\\
\includegraphics[width=80mm,angle=0]{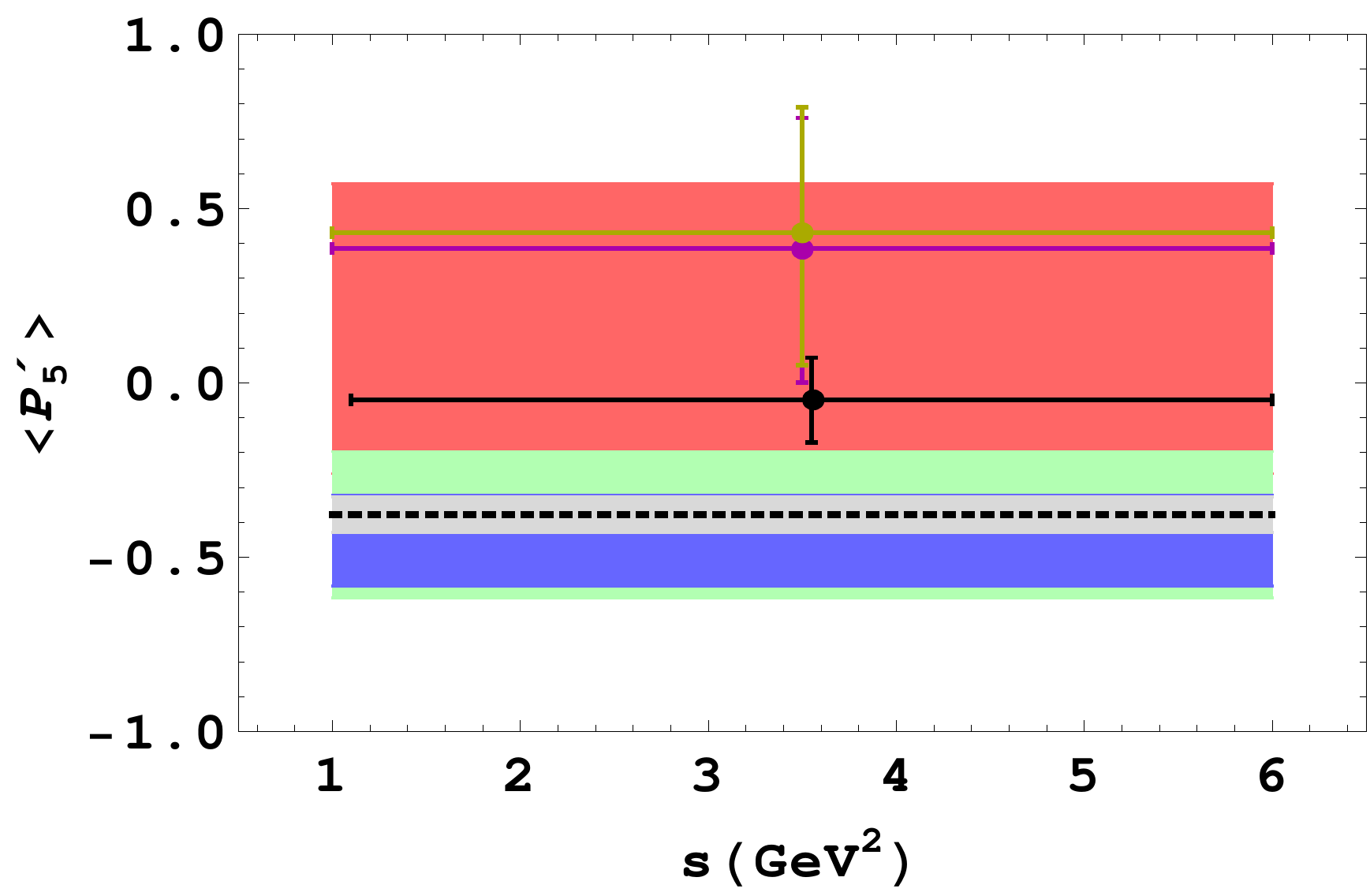}
\includegraphics[width=80mm,angle=0]{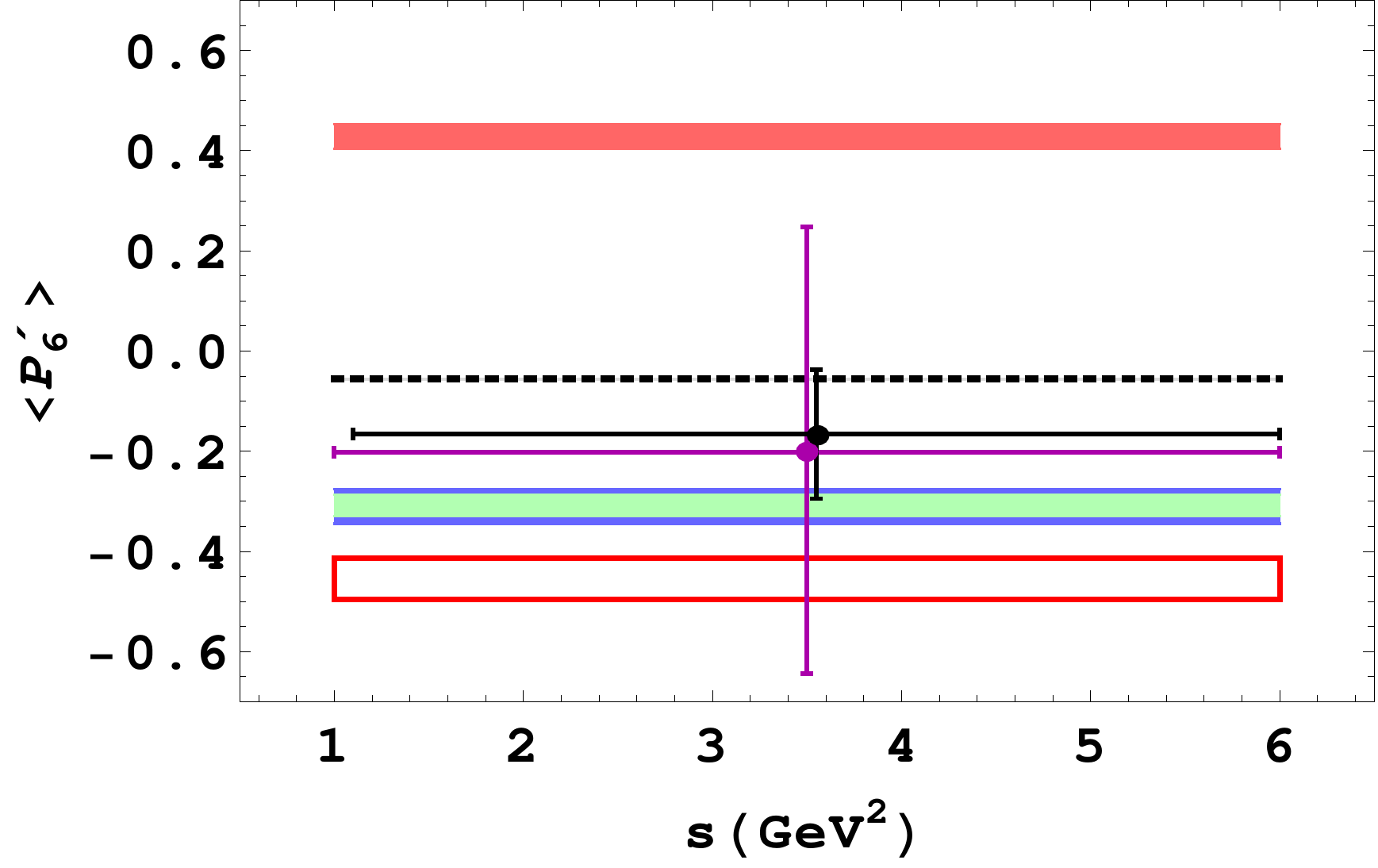}
\end{tabular}
\caption{\sf Optimal observables for $s \in [1.0, 6.0]$ GeV$^2$ where, magenta \cite{belle2016} and
yellow \cite{Wehle:2016yoi} error bars correspond to Belle measurements
available for some of these observables. The empty red box in $\langle P_2 \rangle$ and $\langle P^\prime_6 \rangle$
represents the $\mathcal{S}_3$ when we choose $\phi_{sb} = -150 \pm 10$ given in Tab. \ref{zprimeValues}
of Appendix B. Other legends are same as in Figs. (\ref{p1p}) and (\ref{p2p}).} \label{p12AV16}
\end{center}
\end{figure*}

The four-fold differential decay distribution for the cascade decay
$B\to K^*(\to K\pi)\ell^+\ell^-$ is completely described by the four
independent kinematical variables: the three angles; $\theta_{K^*}$ is the
angle between the $K$ and  $B$ mesons in the rest frame of $K^*$, $\theta_\ell$
is the angle between lepton and $B$ meson in the dilepton rest frame while $\phi$
is the azimuthal angle between the dilepton rest frame and $K^*$ rest frame and the
fourth variable is dilepton invariant squared mass $s$. The explicit dependence of
differential decay distribution on these kinematical variables can be expressed as follows
\begin{align}
\frac{d^4\Gamma}{ds \, d \cos\theta_\ell \, d\cos\theta_{K^*}
  d\phi} &= \frac{9}{32\pi} \widetilde{\Gamma} \left( s,
  \theta_\ell, \theta_{K^*}, \phi \right),\label{diffAngularDis}
\end{align}
where
\footnotesize
\beqa
 &&\widetilde{\Gamma} \left( s,
  \theta_\ell, \theta_{K^*}, \phi \right) = \notag \\
  &&J^s_1 \sin^2\theta_{K^*} + J^c_1 \cos^2\theta_{K^*} + \left(J^s_2 \sin^2\theta_{K^*}
+ J^c_2 \cos^2\theta_{K^*}\right) \cos2\theta_\ell \nn\\
&& + J_3 \sin^2\theta_{K^*} \sin^2\theta_\ell \cos2\phi + J_4 \sin2\theta_{K^*} \sin2\theta_\ell \cos\phi
  \notag \\
  &&+  J_5 \sin2\theta_{K^*} \sin\theta_\ell \cos\phi
+(J_6^s \sin^2\theta_{K^*} +J_6^c \cos^2\theta_{K^*})\cos\theta_\ell\nn \\
 &&+ J_7 \sin2\theta_{K^*} \sin\theta_\ell \sin\phi
+  J_8 \sin2\theta_{K^*} \sin2\theta_\ell \sin\phi \nn\\
&& + J_9 \sin^2\theta_{K^*} \sin^2\theta_\ell \sin2\phi\;.
\eeqa
\normalsize
The full  physical region phase space of kinematical variables is given by
\begin{eqnarray}
&& 4m^2_\ell \leqslant s \leqslant \left(m_B - m_{K^*}\right)^2,\hspace{0.6cm} 0\leqslant \theta_\ell \leqslant \pi,\nn \\
&&  0
\leqslant \theta_{K^*} \leqslant \pi,\hspace{0.7cm} 0\leqslant \phi \leqslant 2\pi,
\end{eqnarray}
where $m_B$, $m_{K^*}$, $m_\ell$ are the masses of $B$ meson, $K^*$ meson and lepton, respectively.

The expressions of  coefficients $J_i^{(a)} = J_i^{(a)}\left(s\right)$ for $i = 1,....,9$ and $a = s,c$ as a function of
the dilepton mass $s$, are given in Appendix A in Eq. (\ref{EXPJ}). As we do not take the scalar contribution in this study, therefore, $J_6^c=0 $.
\subsection{Expressions of the Angular Observables} \label{AnalyticAngObs}

The definitions of FFI angular observables (optimal observables) are given in ref. \cite{Matias:2012xw},
\begin{eqnarray}
P_1(s)&=&\frac{J_3}{2J_2^s},\quad P_2(s)=\beta_\ell\frac{J_6^s}{8J_2^s},\quad P_3(s)=-\frac{J_9}{4J_2^s},\notag \\
P_4(s)&=&\frac{\sqrt{2}J_4}{\sqrt{-J_2^c(2J_2^s-J_3)}},
P_5(s)=\frac{\beta_\ell J_5}{\sqrt{-2J_2^c(2J_2^s+J_3)}}, \notag \\
P_6(s) &=& -\frac{\beta_\ell J_7}{\sqrt{-2J_2^c(2J_2^s-J_3)}}.
\label{Formula1}
\end{eqnarray}
The primed observables (related to the $P_i$ ($i=4,5,6$)) which are simpler and
more efficient to fit experimentally are defined as,
\begin{eqnarray}
P_4^\prime&\equiv&P_4\sqrt{1-P_1}=\frac{J_4}{\sqrt{-J_2^cJ_2^s}},\notag \\
P_5^\prime &\equiv& P_5\sqrt{1+P_1}=\frac{J_5}{2\sqrt{-J_2^cJ_2^s}},\notag \\
P_6^\prime &\equiv& P_6\sqrt{1-P_1}=\frac{-J_7}{2\sqrt{-J_2^cJ_2^s}}.
\label{Formula2}
\end{eqnarray}
\section{Results and Discussion} \label{results}
In this section, we will present the numerical analysis of the angular observables. The authors would like to mention here that all of the numerical results are taken from the self-written {\sf Mathematica} code.
Before the analysis, we would like to write the different definitions of angular observables that are opted by LHCb \cite{2015lhcb} and theoretically used in the literature,
\begin{eqnarray}
P_2^{\text{exp}}&=&-P_2, \quad P_3^{\text{exp}}=-P_3, \quad P_4^{\prime\text{exp}}=-\frac{1}{2}\,P_4^\prime, \quad
\notag \\P_6^{\prime\text{exp}}&=&-P_6^\prime,\quad
P_1^{\text{exp}}= P_1,\quad P_5^{\prime\text{exp}} = P_5^\prime.
\end{eqnarray}

For the numerical analysis the values of LCSR form factors, and relevant fit parameters are listed in Tab. (\ref{TabForm}). The values of Wilson coefficients and other input parameters are listed in Appendix A in Tabs. (\ref{TabInputs}) and (\ref{TabInputs1}), respectively. Regarding the coupling parameters of $Z^\prime$ with quarks and leptons, there are some severe  constraints come from different inclusive and exclusive $B-$ meson channels \cite{ConstrainedZPC1}. Particularly, coming from the two different fitting values for $B_s-\bar B_s$ mixing data by the UTFit collaboration\cite{UTfit}. In this study, we call these two fitting values as $\mathcal{S}_1$ and $\mathcal{S}_2$ and their numerical values are listed in Tab.(\ref{ZP table}). We have considered another scenario which denoted by $\mathcal{S}_3$ in the present study that are obtained from the analysis of $B\to X_{s}\mu^+\mu^-$ \cite{newcon1a}, $B\to K^{*}\mu^+\mu^-$ \cite{newcon1,newcon2} and $B\to \mu^+\mu^-$ \cite{newcon3}. The numerical values of scenario $\mathcal{S}3$ are chosen from \cite{cpv5, newcon4} and also listed in the Tab. (\ref{ZP table}). The purpose of the following analysis is to check that these constrained of $Z^\prime$ parameters could accomodate the anomalies in the angular observables, particularly, in $P_5^\prime$.
\subsection{$P$-observables in different bin size}
The numerical values of angular observables in different low $s$ bins in SM and in $\mathcal{S}_1$, $\mathcal{S}_2$ and $\mathcal{S}_3$ are given in Tab. (\ref{Pbin}). For comparison with experimental measurements, the maximum likelihood fit results of LHCb \cite{2015lhcb} are also given in the table. The ranges in the values of angular observables in $\mathcal{S}_1$, $\mathcal{S}_2$ and $\mathcal{S}_3$ is found by setting the upper and lower values of parametric space of these scenarios.
These results are also shown graphically in Figs. (\ref{p1p}) and (\ref{p2p})
where black crosses are the data points taken from the last column of Tab. (\ref{Pbin}) and black dashed
line correspond to the SM while green, red and blue bands correspond to the
$\mathcal{S}_1$, $\mathcal{S}_2$ and $\mathcal{S}_3$ scenarios of the $Z^\prime$ model, respectively.
The upper curve of the band corresponds the upper values of parametric space while the lower curve of the band
corresponds the lower values of parametric space of the scenario.
 In our different bin sized analysis, we  have not included the preliminary results from
 Belle \cite{belle2016,Wehle:2016yoi},  ATLAS \cite{ATLAS-CONF-2017-023} and CMS\footnotemark \cite{CMS-PAS-BPH-15-008,Khachatryan:2015isa} because their bin intervals are different from LHCb \cite{2015lhcb} that we have discussed in this section.
In Fig. (\ref{p1p}), the gray shaded region corresponds to the uncertainty in the SM values due to the uncertainty in different input parameters. One can see from the left panel of Figs. (\ref{p1p}) and (\ref{p2p})
that the uncertainty band in SM not preclude the effects of $Z^\prime$ model. Therefore, we have not provided the SM
uncertainty in Tab. (\ref{Pbin}) and hence in the right panel of Figs. (\ref{p1p}) and (\ref{p2p}).
\footnotetext{see figure 6 of \cite{Altmannshofer:2017fio} for the recent analysis with these new results.}
\begin{table*}[ht]
\caption{\sf Results for $\langle P \rangle$-observables and their comparison with maximum
likelihood fit results of ref. \cite{2015lhcb} in
different bin size.} \label{Pbin}
\begin{tabular}{|P{2.3cm}|P{2.0cm}|P{3cm}|P{3cm}|P{3.15cm}|P{3.5cm}|}
\hline
{\sf Obs.}           & {\sf SM Prediction} & $\mathcal{S}_1$ & $\mathcal{S}_2$ & $\mathcal{S}_3$ & {\sf Measurement} \cite{2015lhcb} \\
\hline \hline
$0.1<s<0.98$ GeV$^2$       &  &  &  &  &    \\
$\langle P_1\rangle$        & $-0.002$  &  $-0.002\leftrightarrow-0.008$  &  $-0.002\leftrightarrow-0.002$ &  $-0.002\leftrightarrow-0.009$ &  $-0.099^{+0.168}_{-0.163}\pm0.014 $    \\
$\langle P_2\rangle$        & $-0.106$ & $-0.134\leftrightarrow-0.113$ & $-0.116\leftrightarrow-0.102$  & $0.042\leftrightarrow-0.059$  & $-0.003^{+0.051}_{-0.052}\pm0.007 $  \\
$\langle P_3\rangle$        & $-0.0001$ &  $0.000\leftrightarrow-0.0002$ &  $-0.000\leftrightarrow-0.001$ &  $-0.000\leftrightarrow-0.0001$ & $0.113^{+0.079}_{-0.079}\pm0.006 $ \\
$\langle P_4^\prime\rangle$  & $0.267$ &  $0.175\leftrightarrow0.155$ &  $0.230\leftrightarrow0.171$  &  $0.405\leftrightarrow0.380$ &  $0.185^{+0.158}_{-0.154}\pm0.023$ \\
$\langle P_5^\prime\rangle$  & $0.740$  & $0.747 \leftrightarrow 0.473$ & $0.712 \leftrightarrow 0.497$ &$-0.209 \leftrightarrow 0.424$ &$0.387^{+0.132}_{-0.133} \pm 0.052$ \\
$\langle P_6^\prime\rangle$  & $-0.158$ &  $-0.447\leftrightarrow-0.585$  &  $-0.384\leftrightarrow-0.566$  &  $0.466\leftrightarrow0.400$ & $0.034^{+0.134}_{-0.135}\pm0.015$ \\
\hline \hline
$1.1<s<2.5$ GeV$^2$       &  &  &  &  &   \\
$\langle P_1\rangle$           & $-0.007$ &  $-0.008\leftrightarrow-0.008$ &  $-0.007\leftrightarrow-0.008$ &  $-0.006\leftrightarrow-0.006$  & $-0.415^{+0.519}_{-0.636}\pm0.038 $  \\
$\langle P_2\rangle$           & $-0.433$ & $-0.417\leftrightarrow-0.161$  & $-0.406\leftrightarrow-0.187$  & $0.097\leftrightarrow-0.347$ &  $-0.373^{+0.146}_{-0.199}\pm0.027 $   \\
$\langle P_3\rangle$     & $0.0001$ &  $-0.000\leftrightarrow0.001$ &  $0.000\leftrightarrow0.001$ &  $0.001\leftrightarrow0.001$  &  $0.350^{+0.330}_{-0.254}\pm0.015 $ \\
$\langle P_4^\prime\rangle$    & $0.023$ &  $-0.113\leftrightarrow-0.173$ &  $-0.040\leftrightarrow-0.13$ &  $0.170\leftrightarrow0.200$  &$-0.163^{+0.232}_{-0.240}\pm0.021$ \\
$\langle P_5^\prime\rangle$    &$0.225$ & $0.275  \leftrightarrow -0.208$  & $0.211\leftrightarrow -0.141$  & $0.249\leftrightarrow 0.160$ & $0.289^{+0.220}_{-0.202}\pm0.023$ \\
$\langle P_6^\prime\rangle$    & $-0.078$ &  $-0.432\leftrightarrow-0.533$  &  $-0.400\leftrightarrow-0.536$ &  $0.689\leftrightarrow0.520$ & $-0.463^{+0.202}_{-0.221}\pm0.012$ \\
\hline \hline
$2.5<s<4.0$ GeV$^2$       &  &  &  &  &    \\
$\langle P_1\rangle$           & -0.023 &  $-0.025\leftrightarrow-0.026$  &  $-0.024\leftrightarrow-0.025$ &  $-0.032\leftrightarrow-0.024$ &  $0.571^{+2.404}_{-1.714}\pm0.045 $  \\
$\langle P_2\rangle$           & $-0.228$ & $-0.215\leftrightarrow0.154$ & $-0.188\leftrightarrow0.110$ & $-0.341\leftrightarrow-0.280$  &  $-0.636^{+0.444}_{-1.735}\pm0.015 $   \\
$\langle P_3\rangle$     & 0.001 &  $-0.000\leftrightarrow0.002$  &  $0.001\leftrightarrow0.002$  &  $0.004\leftrightarrow0.004$ &  $0.745^{+2.587}_{-0.861}\pm0.030 $ \\
$\langle P_4^\prime\rangle$    & $-0.282$ &  $-0.355\leftrightarrow-0.394$ &  $-0.320\leftrightarrow-0.371$  &  $-0.314\leftrightarrow-0.205$  &  $-0.713^{+0.410}_{-1.305}\pm0.024$ \\
$\langle P_5^\prime\rangle$    & $-0.400$ & $-0.204\leftrightarrow -0.667$ & $-0.339\leftrightarrow -0.628$& $0.722\leftrightarrow -0.294$  & $-0.066^{+0.343}_{-0.364}\pm0.023$ \\
$\langle P_6^\prime\rangle$    & $-0.066$ &  $-0.313\leftrightarrow-0.350$ &  $-0.309\leftrightarrow-0.372$ &  $0.568\leftrightarrow0.508$ &  $0.205^{+0.962}_{-0.341}\pm0.013 $ \\
\hline \hline
$4.0 <s< 6.0$ GeV$^2$       &  &  &  &  &     \\
$\langle P_1\rangle$           & -0.055 &  $-0.053\leftrightarrow-0.053$  &  $-0.054\leftrightarrow-0.053$  &  $-0.064\leftrightarrow-0.062$  & $0.180^{+0.364}_{-0.348}\pm0.0.027 $   \\
$\langle P_2\rangle$           & $0.206$ & $0.088\leftrightarrow0.357$  & $0.170\leftrightarrow0.341$ & $-0.407\leftrightarrow0.146$  &  $0.042^{+0.088}_{-0.087}\pm0.011 $  \\
$\langle P_3\rangle$     & $0.001$ &  $-0.000\leftrightarrow0.003$  &  $0.000\leftrightarrow0.003$ &  $0.003\leftrightarrow0.004$ &  $0.083^{+0.187}_{-0.184}\pm0.023 $ \\
$\langle P_4^\prime\rangle$    & $-0.443$  &  $-0.460\leftrightarrow-0.472$ &  $-0.452\leftrightarrow-0.465$ &  $-0.477\leftrightarrow-0.446$  & $-0.448^{+0.169}_{-0.172}\pm0.020$ \\
$\langle P_5^\prime\rangle$    & $-0.761$ & $-0.492\leftrightarrow -0.837$ & $-0.653\leftrightarrow-0.829$ & $0.682\leftrightarrow -0.514$ & $-0.300^{+0.158}_{-0.159}\pm0.023$ \\
$\langle P_6^\prime\rangle$    & $-0.036$  &  $-0.182\leftrightarrow-0.198$ &  $-0.178\leftrightarrow-0.214$  &  $0.249\leftrightarrow0.268$ &  $-0.032^{+0.167}_{-0.166}\pm0.007 $\\
\hline
\end{tabular}
\end{table*}

The plots in first and third rows of Fig. (\ref{p1p}), represent the variation in the
values of $P_{1,3}$ and their average values $\langle P_{1,3}\rangle$ as a function
of $s$ in the SM and in the different scenarios of $Z^\prime$ model.
From these graphs one can see that the values of these observables are quite small
in the SM and not much enhanced when we incorporate the $Z^\prime$ effects.
One can also see from Fig. (\ref{p1p}) that the SM values of $\langle P_1\rangle$ lie
inside the measured values. As the error in the measurement is huge, therefore, no potent result can
be drawn from this observable with the current data. On the other hand the values of
$\langle P_3\rangle$ in last two bins are within the measured values while in first
two bins the SM values are out of the measured bars. However, to say something about
any discrepancy in these observables, reduction in the experimental  uncertainties are required.

Plots in second row of Fig. (\ref{p1p}), show the variation in the values of
$P_2$ and its average $\langle P_2\rangle$ against dilepton mass $s$.
It could be seen from these figures that the values of these observables are
significantly influenced in the presence of $Z^\prime$ effects. The right
plot in the second row of Fig. (\ref{p1p}) shows that the SM values of
$\langle P_2\rangle$  in the bins $s\in[1.1,2.5]$ and $s\in[2.5,4.0]$ lie
within the measurements and also in the bin $s \in[4.0,6,0]$ when the
theoretical  uncertainties of the input parameters are taken into account.
However, in the first bin $s\in[0.1,0.98]$, the SM value of  $\langle P_2\rangle$
looks mismatch from the experimental value. But it is worthy to mention here that
the measurement performed by LHCb in this bin is without including the $m_\ell-$
suppressed terms which are important at very low $s$ region and  it was found
in \cite{Descotes-Genon:2015uva}, that the impact of these terms is about
$23\%$ reduction in the value of $\langle P_2\rangle$. Regarding this, it is
mentioned in \cite{Capdevila:2017ert} that in the first bin,
LHCb actually measured $\langle\hat{ P}_2\rangle$ instead of $\langle P_2\rangle$.
Therefore, in principle, one could say that, up-till now, there is no mismatch
between the SM predicted values  of $\langle P_2\rangle$ with the  experimental values.

In the first row of Fig. (\ref{p2p}), we have displayed  $P_4^\prime$ and it's
average value, $\langle P_4^\prime\rangle$, in the SM and in the different scenarios of
$Z^\prime$ model as a function of $s$. One can see from these plots that the $Z^\prime$
effects are quite significant in the  $P_4^\prime$ values at low $s$ region but mild
at larger values of $s$. However, the SM values of
$\langle\mathcal {P}_4^\prime\rangle$ in all four bins lie inside the measured values.

The results of $P_5^\prime$ and it's average value $\langle P_5^\prime\rangle$
in the SM and in the $Z^\prime$ models are presented in the second row of
Figs. (\ref{p2p}). The values are significantly changed from the SM values when we
incorporate the $Z^\prime$ effects. It can be noticed in the bin $s=4\text{ to }6$ GeV$^2$,
the SM average value $\langle P_5^\prime\rangle$ mismatch with the experimental values
and as mentioned in the introduction that LHCb found 3$\sigma$ deviation in this bin.
It could be seen from the figure that this discrepancy can be alleviated by $\mathcal{S}_3$ (red band) of $Z^\prime$ model.
On the other hand for the Utfit scenarios, namely, $\mathcal{S}_1$ and $\mathcal{S}_2$ it can be noticed that when we take the upper and lower limit values of the current parametric space of these scenarios (green and blue bands), the $P_5^\prime$ anomaly in the  bin $s\in[4,6]$GeV$^2$ can not be accommodated. However, if the values of different parameters are chosen randomly within the allowed range then one could accommodate the $P_5^\prime$ anomaly in this bin by $\mathcal{S}_1$ but not with $\mathcal{S}_2$. Therefore, it looks that the $\mathcal{S}_2$ of Utfit is not consistent with the present data while the parametric space of $\mathcal{S}_1$, the left (right) couplings, ($B_{\ell\ell}^L$, $B_{\ell\ell}^R$), of $Z^\prime$ with leptons is severely constraint as shown in Fig. (\ref{constCou}).

In the third row of Fig. (\ref{p2p}), we have shown the variation of
$P_6^\prime$ and $\langle P_6^\prime\rangle$ as a function of $s$. Similar to $P_{1,3}$, the
SM value of this observable is also suppressed. As seen from the graph that
SM value of $P_6^\prime$ consistent with the data with large error bars, however
there is 2$\sigma$ deviation in one bin $s\in[1.1,2.5]$ which probably will be
disappear when data will increase.
One can also notice that in contrast to the $P_{1,3}$, the value of
$P_6^\prime$ significantly enhanced in the $Z^\prime$ model. It is also
noticed that in the $Z^\prime$ model the value of  $P_6^\prime$ is
positive in scenarios $\mathcal{S}_1$ and $\mathcal{S}_2$ while becomes negative in $\mathcal{S}_3$.
As for the present analysis in $\mathcal{S}_3$, we set the value of $\phi_{sb}=150\pm10$, in contrast to this, if we choose $\phi_{sb}=-150\pm10$ which is also allowed (see Tab. (V)), then this negative value becomes positive.

\subsection{$P$-observables in $s\in[1.0, 6.0]$ GeV$^2$}
Besides the analysis of angular observables in shorter bins at low $s$ region (discussed in previous section), we have also analyzed these observables in the full $s\in[1.0, 6.0]$ GeV$^2$ region. The results for $P$-observables in $s\in[1.0, 6.0]$ GeV$^2$ are summarized in Tab. (\ref{Pbin16})
and corresponding plots are shown in Fig. (\ref{p12AV16}). In this figure, black error bar corresponds to LHCb result \cite{2015lhcb}  while, magenta  and
yellow error bars correspond to Belle measurements
for some of these observables \cite{belle2016,Wehle:2016yoi}. However, it is good to mention here that LHCb results are in the bin $s\in[1.1, 6.0]$ GeV$^2$
while Belle measurements \cite{belle2016,Wehle:2016yoi} are in the $s\in[1.0, 6.0]$ GeV$^2$. In addition, recently, the ATLAS collaboration announced its results for
$s\in[0.04, 6.0]$ GeV$^2$ \cite{ATLAS-CONF-2017-023} which is not included in the current analysis.
The empty red boxes in the plots of $\langle P_2 \rangle$ and $\langle P^\prime_6 \rangle$
represent the $\mathcal{S}_3$ scenario when we choose $\phi_{sb} = -150 \pm 10$, while, other legends are same as in Figs. (\ref{p1p}) and (\ref{p2p}).

From  Fig. (\ref{p12AV16}), one can immediately notice that the values of  $\langle P_1 \rangle$ and $\langle P_4^\prime \rangle$ in the SM and in all the three scenarios of $Z^\prime$ lie within the current measurements, however, the error bars are huge. Therefore, to extract any information about the NP requires the precise measurement of these observables. It is also noticed that the values of  $\langle P_1 \rangle$ in the SM and in the $Z^\prime$ scenarios are very close, consequently, this observable even after the reduction of error bars not a good candidate to constrained the $Z^\prime$ parametric space. On the other hand  $\langle P_4^\prime \rangle$ could be helpful  to constraint the $Z^\prime$ parametric space, if any mismatch will appear in future in the bin [1,6] GeV$^2$. The SM value of $\langle P_3 \rangle$ is small  and not enhanced in $Z^\prime$ model. However, the measured value is well above the SM prediction with huge error bars and need precision to draw any conclusion from this observable as well.
From graph of $\langle P_2 \rangle$ in Fig. (\ref{p12AV16}), one can deduced that the SM value of $\langle P_2 \rangle$ not lie within the measured value of LHCb. However, the values of  $\langle P_2 \rangle$  in $\mathcal{S}_1$ and $\mathcal{S}_2$ are within the measurements while in $\mathcal{S}_3$, the value is out side the measured error bars. For $\langle P^\prime_6 \rangle$, we have two different measurements as shown in the plot and contrast to the $\langle P_2 \rangle$, the value of $\langle P^\prime_6 \rangle$ lie within these measurements. However, similar to $\langle P_2 \rangle$ the values of $\langle P^\prime_6 \rangle$ in $\mathcal{S}_1$ and $\mathcal{S}_2$ lie within the measurements while the value in $\mathcal{S}_3$ lies outside the measured values (see red bands in both plots). Regarding $\mathcal{S}_3$, it is interesting to check whether the values of $\langle P_2 \rangle$ and $\langle P^\prime_6 \rangle$ could be reduced  to current  measurements. For this purpose,  we choose the weak phase with opposite sign i.e., $\phi_{sb}=-150\pm10$ (see Tab. \ref{zprimeValues} in Appendix A) and represent them in plots by empty red boxes. In Fig. 4, by looking the empty red box in $\langle P_2 \rangle$ plot, the value is reduced but still well above the current measurement. In contrast, the value of $\langle P^\prime_6 \rangle$ reduce to
the Belle measurements \cite{belle2016}. However, more statistics on the observables  $\langle P_2 \rangle$ and  $\langle P^\prime_6 \rangle$ are helpful to constrained the $Z^\prime$ parameters, particularly, the sign and the magnitude of  new weak phase $\phi_{sb}$.
\begin{figure*}[th]
\begin{center}
\begin{tabular}{lcr}
\includegraphics[width=80mm,angle=0]{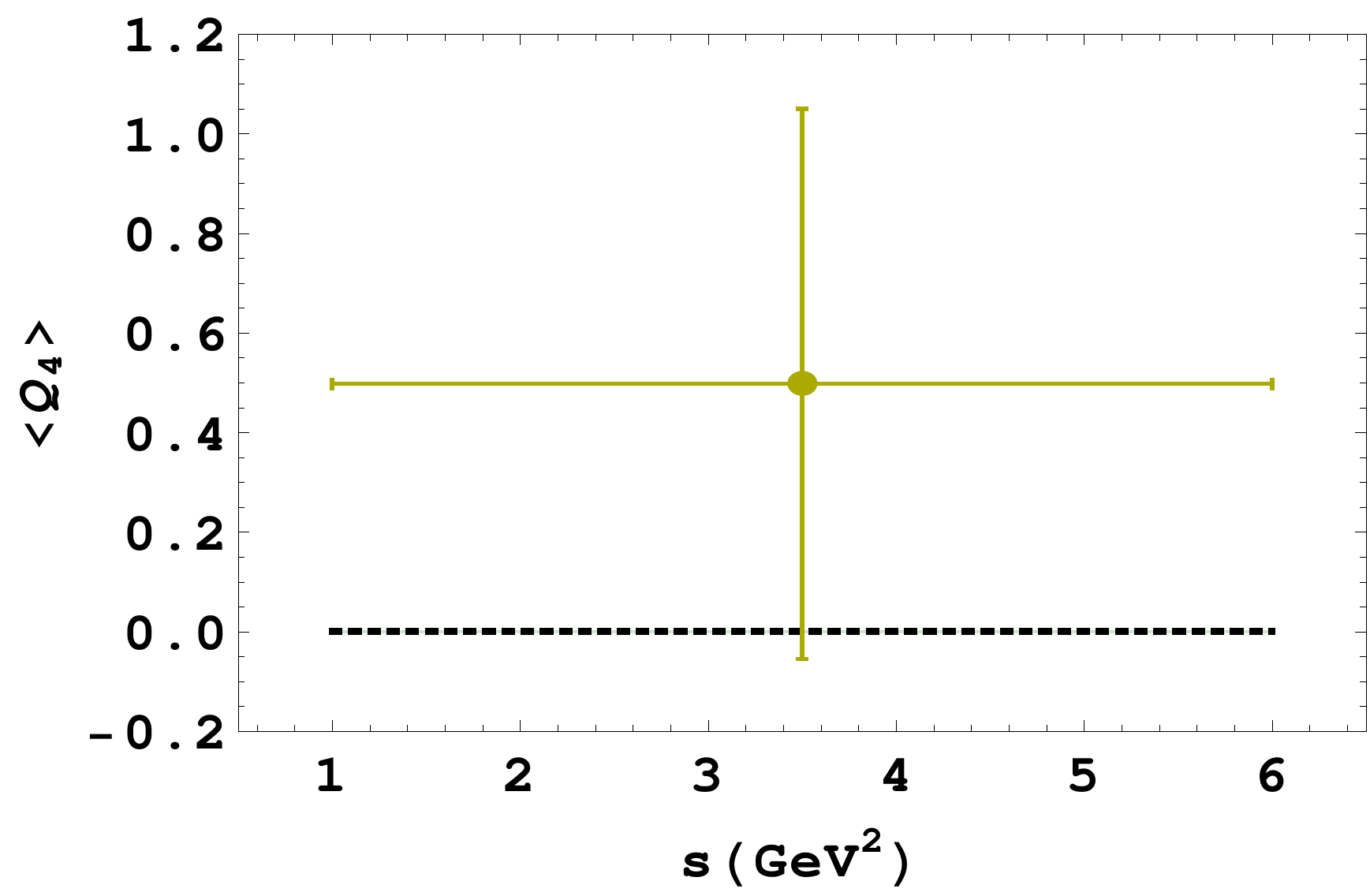}
\includegraphics[width=80mm,angle=0]{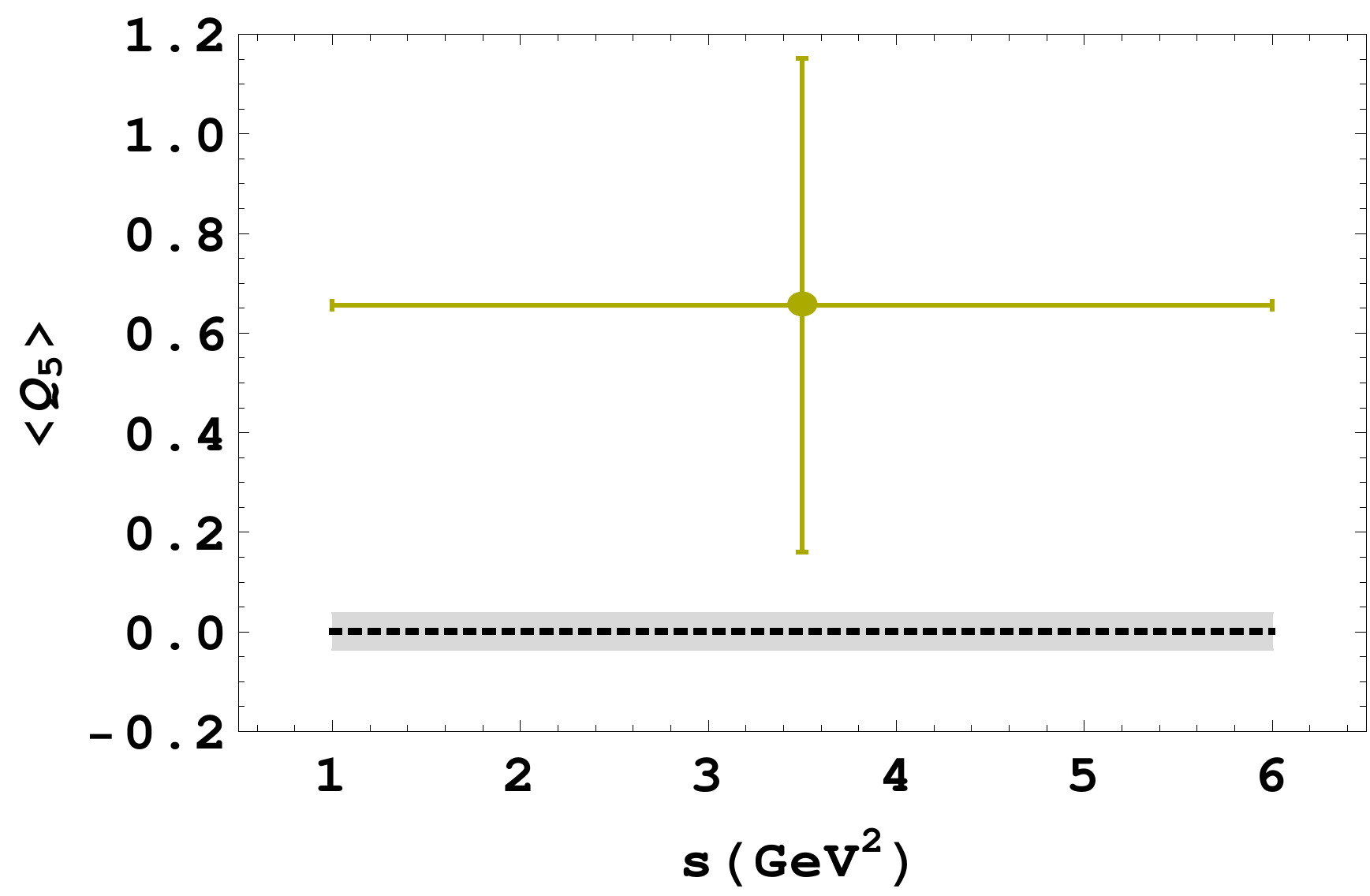}
\end{tabular}
\caption{\sf Optimal observables $Q_4$, $Q_5$ for $s \in [1.00,6.00]$
GeV$^2$ where, yellow error bar corresponds to recent Belle measurements \cite{Wehle:2016yoi}.
Other legends are same as in previous figures.} \label{q4q5}
\end{center}
\end{figure*}

For  $\langle P^\prime_5 \rangle$ plots of Fig. (\ref{Pbin16}), the values in the SM and in $\mathcal{S}_1$, $\mathcal{S}_2$ lie out side the error bars of experimental data points while the values in the $\mathcal{S}_3$  well inside the all data points shown in figure. In general, from the plots of Fig. (\ref{Pbin16}), one concludes that the considered model do have potential to remove mismatch
between theory and experiment but it is not so conclusive at present. We hope more precise measurements will clear the
situation.
\begin{table*}[ht]
\caption{\sf Results for $\langle P\rangle$-observables for $s\in[1.0, 6.0]$ GeV$^2$ and their comparison
with LHCb maximum likelihood fit results of ref. \cite{2015lhcb} in
different bin size, Belle results \cite{belle2016,Wehle:2016yoi}.} \label{Pbin16}
\begin{center}
\begin{tabular}{|P{1.4cm}|P{2.4cm}|P{3cm}|P{3cm}|P{3cm}|P{3.8cm}|}
\hline
{\sf Obs.}           & {\sf SM Prediction} & $\mathcal{S}_1$ & $\mathcal{S}_2$ & $\mathcal{S}_3$ & {\sf Measurement}  \\
\hline \hline
$\langle P_1\rangle$           & -0.033$\pm$0.001 & $-0.032\leftrightarrow-0.034$  & $-0.033\leftrightarrow-0.033$   &$-0.039\leftrightarrow-0.036$   & $0.080^{+0.248}_{-0.245}\pm0.044$\,\cite{2015lhcb}  \\
$\langle P_2\rangle$           & $0.091\pm0.033$  & $0.133\leftrightarrow-0.162$   & $0.087\leftrightarrow-0.135$  & $0.254\leftrightarrow0.106$  & $-0.162^{+0.072}_{-0.073}\pm0.010$\,\cite{2015lhcb}  \\
$\langle P_3\rangle$     & $0.001\pm0.000$  & $-0.000\leftrightarrow0.002$  & $0.000\leftrightarrow0.002$  & $0.003\leftrightarrow0.003$  & $0.205^{+0.135}_{-0.134}\pm0.017$\,\cite{2015lhcb} \\
$\langle P_4^\prime\rangle$    & $-0.264\pm0.014$ & $-0.333\leftrightarrow-0.368$  & $-0.298\leftrightarrow-0.347$  &  $-0.195\leftrightarrow-0.256$ & $-0.336^{+0.124}_{-0.122}\pm0.12$\,\cite{2015lhcb}  $-0.095^{+0.302}_{-0.309}\pm0.174$\,\cite{belle2016} $-0.22^{+0.35}_{-0.34}\pm0.15$\,\cite{Wehle:2016yoi} \\
$\langle P_5^\prime\rangle$    & $-0.378\pm0.051$ & $-0.197\leftrightarrow-0.617$ & $ -0.322\leftrightarrow-0.583$ & $0.572\leftrightarrow-0.260$ &   $-0.049^{+0.107}_{-0.108}\pm0.014$\,\cite{2015lhcb} $0.385^{+0.276}_{-0.285}\pm0.099$\,\cite{belle2016} $0.43^{+0.26}_{-0.28}\pm0.10$\,\cite{Wehle:2016yoi} \\
$\langle P_6^\prime\rangle$    & $-0.056\pm-0.000$ &  $ -0.287\leftrightarrow-0.330$  &  $ -0.276\leftrightarrow-0.345$ &  $ 0.452\leftrightarrow0.403$ &  $-0.166^{+0.108}_{-0.108}\pm0.021$\,\cite{2015lhcb}  $-0.202^{+0.278}_{-0.270}\pm0.172$\,
\cite{belle2016} \\
\hline
\end{tabular}
\end{center}
\end{table*}

\subsection{$Q_{4,5}$ for $ s \in [1.0,6.0]$ GeV$^2$}
In Fig. (\ref{q4q5}), we have plotted the lepton flavor universality violation (LFUV) observables
$\langle Q_{4(5)}\rangle$ against $s$. The values are quite small in the SM
approximately $\langle Q_{4(5)}\rangle=8.8\pm2.1\times10^{-3}(7.5\pm3.6\times10^{-3})$
in the bin $s\in[1,6]$GeV$^2$. We have also found  that the effects of $Z^\prime$ are
negligible. This fact is trivial, since Eq. (3) implies $C_{9,10}^{Z^\prime,\mu}=C_{9,10}^{Z^\prime,e}$. However, error bars  are  quite  large and need more experimental data to find the accurate values of these observables.

%
\section{ Conclusion}
In the present study, we have calculated the angular observables $P_i$
and their average values $\langle P_i\rangle$ in the SM and in the noun-universal family of $Z^\prime$
model for the decay channel $B\to K^*(\to K \pi)\ell^+\ell^-$. The expressions of the angular observables
are given in the form of coefficient $J_i(s)$ which are written in terms of auxiliary
functions $g_i(h_i)$ in Eq. (\ref{EXPJ}). As in the literature, these $J_i(s)$ coefficients, in general,
expressed via transversity amplitudes, $A_\perp$, $A_\parallel$ and $A_0$, so the
relations of these transversity amplitudes with auxiliary function $g_i(h_i)$ are also given in Eq. (\ref{TARelation}).
To see the $Z^\prime$ effects on these observables, we have used the UtFit collaboration constraints for the
$Z^\prime$ parameters, called as scenarios $\mathcal{S}_1$ and $\mathcal{S}_2$.
Besides, we also consider another scenario, called $\mathcal{S}_3$, in the present study.
 From the present analysis, in all three scenarios of $Z^\prime$ for small values of $s$ i.e the large recoil region,
the values of angular observables are
significantly changed from their SM values.
The current analysis shown that except the sceanario $\mathcal{S}_1$, the scenarios $\mathcal{S}_2$ and $\mathcal{S}_3$ of $Z^\prime$ model has potential to
accommodate the mismatch between the recent experimental measurements and the SM values of
some of the angular observables in some bins of $s$.
For instance, there is a discrepancy between experimentally
measured value and SM value of $P_5^\prime$ in the region $s\in[4,6]$ GeV$^2$ and in the current study it is found that
scenario $\mathcal{S}_3$ of $Z^\prime$ could be adjusted this mismatch value with
the measured value in this bin. On the other hand, this mismatch can not be accommodated on taking the maximum and minimum values of different parameters of scenarios $\mathcal{S}_1$ and $\mathcal{S}_2$ of UtFit collaborations.
  However, when we choose the random values of different parameters in the allowed region of these scenarios, one can accommodate the $P_5^\prime$ anomaly with scenario $\mathcal{S}_2$ but not with scenario $\mathcal{S}_1$. It is also noticed that the $P_5^\prime$ anomaly further constraint on the allowed parametric space of $\mathcal{S}_2$. Furthermore, we have also calculated the angular observables $\langle P_i\rangle$
 and the LFUV observables $\langle Q_{4,5}\rangle$ in the large bin $s\in[1,6]$ and plotted with the measured data, however,
 the error bar is quite large in this bin and more static is needed to draw results.
Here, we would like to comment that CMS and ATLAS collaborations recently announced preliminary results on angular observables
in Moriond 2017 which still show the tension between experimental measurements and the SM predictions.
Therefore, in general, one can say,
as data will be enlarge and the
statistical error will be reduced then these observables are quite promising
to say something about the constraints on coupling of $Z^\prime$ boson with
the quarks and leptons and consequently about the status of $Z^\prime$ model.

\appendix
\section{}
The expressions of $J_i$ appeared in Eqs. (\ref{Formula1}) and (\ref{Formula2}) are as follows,
\begin{eqnarray}
J_1^s&=&\frac{3s\beta  _\ell^2}{2} \bigg[p_{K^*}^2 s\left(|g_1|^2+|h_1|^2\right)
   +|g_2|^2 +|h_2|^2\bigg]\notag \\
   &&+\frac{8 m _\ell^2}{s}(p_{K^*}^2 s|h_1|^2 +|h_2|^2), \notag \\
J_1^c&=&\frac{2}{m_{K^*}^2} \bigg[ 32\, \text{a}_0^2 \,C_{10}^{\rm tot^ 2} m_{K^*}^2 m _\ell^2p_{K^*}^2+\beta  _\ell^2s
  |E_{K^*} g_2\notag \\ &+&2\sqrt{s}p_{K^*}^2 g_3|^2+ \left(2-\beta  _\ell^2\right)s
  |E_{K^*} h_2+2\sqrt{s}p_{K^*}^2 h_3|^2\bigg],\notag \\
J_2^s&=&\frac{1}{2} s \beta  _\ell^2 \bigg[P_{K^*}^2 s
   \left(|g_1|^2+|h_1|^2\right)+|g_2|^2+|h_2|^2\bigg],\notag \\
J_2^c&=&-\frac{2 \beta  _\ell^2s}{m_{K^*}^2} \bigg[
  |E_{K^*} \text{g}_2+2\sqrt{s}p_{K^*}^2 g_3|^2
  +|E_{K^*} h_2\notag \\
  &&+2\sqrt{s}p_{K^*}^2 h_3|^2\bigg], \notag
\end{eqnarray}
\begin{eqnarray}
J_3&=&s \beta  _\ell^2 \bigg[p_{K^*}^2 s
   \left(|g_1|^2+|h_1|^2\right)-|g_2|^2-|h_2|^2\bigg],\notag \\
J_4&=&\frac{\sqrt{2}s\beta  _\ell^2}{m_{K^*}}  \bigg[E_{K^*}  \left(|g_2|^2+|h_2|^2\right)\notag \\
&&\qquad+p_{K^*}^2
   \left(s\right)^{1/2} 2\mathcal{R}e( g_2g_3^*+ h_2h_3^*)\bigg],\notag \\
J_5&=&-\frac{\sqrt{8}p_{K^*}\left(s\right)^{3/2} \beta  _\ell}{m_{K^*}} \bigg[ E_{K^*} \mathcal{R}e(g_1h_2^* +  g_2h_1^*)\notag \\ &&\qquad+2p_{K^*}^2 s^{1/2}\mathcal{R}e( g_1 h_3^*+ g_3h_1^* )\bigg],\notag \\
J_6^s&=&-4p_{K^*} \left(s\right)^{3/2} \beta  _\ell \bigg[ \mathcal{R}e(g_1 h_2^*+g_2 h_1^*)\bigg],\notag \\
J_7 &=& \frac{ \sqrt{32}p_{K^*}^2 \left(s\right)^{3/2} \beta  _\ell}{m_{K^*}} \bigg[\mathcal{I}m(g_2 h_3^*+g_3^*h_2)\bigg],\notag \end{eqnarray}
\begin{eqnarray}
J_8&=&\frac{\sqrt{2}p_{K^*}\left(s\right)^{3/2} \beta  _\ell^2}{ m_{K^*}} \bigg[E_{K^*}\mathcal{I}m(g_1^*g_2+ h_1^*h_2)\notag \\ &&\qquad+2p_{K^*}^2s^{1/2}\mathcal{I}m(g_1^*g_3+h_1^*h_3)\bigg],\notag \\
J_9&=& 2p_{K^*} \left(s\right)^{3/2} \beta  _\ell^2\bigg[ \mathcal{I}m(g_1g_2^*+h_2h_1^*)\bigg],\label{EXPJ}
\end{eqnarray}
where $g_i$($h_i$), $i=1,\cdots,3$ are the auxiliary functions and given as follows,
\begin{eqnarray}
h_1&=&\frac{4  m_b}{s} \mathcal{T}_\perp+\frac{2 }{M_B+m_{K^*}}C_9^{\rm tot} V(s), \notag \\
 g_1 &=& \frac{2 }{M_B+m_{K^*}}C_{10}^{\rm tot} V(s), \quad \notag \\
h_2& =& - (M_B+m_{K^*}) C_9^{\rm tot}A_1(s)\notag \\
&&-\frac{4  m_b\left(m_B^2-m_{K^*}^2\right)}{s}\frac{E_{K^*}}{M_B}\mathcal{T}_\perp,\notag \\
g_2&=&- (M_B+m_{K^*})A_1(s) C_{10}^{\rm tot}, \notag \\
h_3&=&\frac{A_2 }{M_B+m_{K^*}}C_9^{\rm tot}\notag \\
&&+\frac{2 m_b}{s} \bigg[\frac{s(\mathcal{T}_\perp+\mathcal{T}_\parallel)
}{m_B^2-m_{K^*}^2}+\frac{2E_{K^*}}{M_B}\mathcal{T}_\perp\bigg], \notag \\
g_3&=&\frac{A_2 }{M_B+m_{K^*}}C_{10}^{\rm tot} \label{axfu},
\end{eqnarray}
\begin{eqnarray}
E_{K^*}&=&\frac{m_B^2-m_{K^*}^2-s}{2 \sqrt{s}}, \quad p_{K^*} = \sqrt{E_{K^*}^2-m_{K^*}^2}, \notag \\
 \beta_\ell&=&\sqrt{1-\frac{4m^2_\ell}{s}}\label{EKdef},
\end{eqnarray}
and a$_0=\frac{E_{K^*}}{m_{K^*}}\frac{\xi_{\parallel}}{\Delta_{\parallel}}$ where $\Delta_{\parallel}$ is given in Appendix B in Eq. (\ref{delta}).

Traditionally, the $J$'s are given in terms of transversity amplitudes but we have written in terms of $g_i(h_i)$ functions given in Eq. (\ref{axfu})
$A_{0,\parallel,\perp}$. The $A_{0,\parallel,\perp}$  are related with $g_i(h_i)$  as follows
\begin{eqnarray}
A_0^{L,R}&=&\frac{\mathcal{N}}{m_{K^*}} \bigg[E_{k^*}  (h_2\mp g_2)+2 p_{k^*}^2 \sqrt{s} (h_3\mp g_3)\bigg], \notag \\
A_\parallel^{L,R}&=&\sqrt{2}\mathcal{N} \big[  h_2\mp g_2\big],\quad
A_\perp^{L,R}=\sqrt{2s}\mathcal{N}p_{k^*} \big[h_1\mp g_1\big],\notag \\
\label{TARelation}
\end{eqnarray}
where $\mathcal{N}=\alpha \, G_F |V_{tb}V_{ts}^*|\sqrt{\frac{s\,\beta_\ell\, p_{K^*}}{3\cdot2^{10}\pi^5m_B^3}}$. We would like to mention here that our expressions of $J$'s are consistent with the literature for example given in refs. \cite{Matias:2012xw,ball}. 

The values of Wilson coefficients at NNLO, $Z^\prime$ parameters and other input parameters are listed in Tabs. (\ref{TabInputs}),  (\ref{ZP table}) and (\ref{TabInputs1}), respectively.
\begin{table*}[ht]
\centering \caption{ \sf Values of Wilson coefficients at $\mu_b=4 \cdot 8$.}
\begin{tabular}{|cccccccccc|}
\hline
$C_1(\mu_b)$ &   $C_2(\mu_b)$ &  $C_3(\mu_b)$ &  $C_4(\mu_b)$
& $C_5(\mu_b)$ & $C_6(\mu_b)$ & $C_7^{\rm eff}(\mu_b)$ & $C_8^{\rm eff}(\mu_b)$
& $C_9(\mu_b)$ & $C_{10}(\mu_b)$ \\
\hline
-0.2632 & 1.0111 & -0.0055 & -0.0806 & 0.0004 &
0.0009 &  -0.2923 & -0.1663 & 4.0749 & -4.3085\\
\hline\hline
\end{tabular}
\label{TabInputs}
\end{table*}
\begin{table*}[ht]
\centering
\caption{\sf The numerical values of the $Z^\prime$ parameters
\cite{ConstrainedZPC1,UTfit
}.}\label{zprimeValues}
\begin{tabular}{|c|cccc|c|}
\hline
 &  $|\mathcal{B}_{sb}|\times10^{-3}$  & $\phi_{sb}(\text{Degree})$  &  $S_{\ell\ell}^{LR}\times10^{-2}$   & $D_{\ell\ell}^{LR}\times10^{-2}$\\
\hline
 \ \  $\mathcal{S}1$ \ \  & \ \  $1.09\pm0.22$ \ \  & \ \  $-72\pm7$ \ \  & \ \  $-2.8\pm3.9$ \ \  & \ \  $-6.7\pm2.6$ \\
\ \  $\mathcal{S}2$ \ \  & \ \  $2.20\pm0.15$ \ \  & \ \  $-82\pm4$ \ \  & \ \  $-1.2\pm1.4$ \ \  & \ \  $-2.5\pm0.9$  \\
\ \  $\mathcal{S}3$ \ \  & \ \  $4.0\pm1.5$ \ \  & \ \  $150\pm10$ or $(-150\pm10)$ \ \  & \ \  $0.8$ \ \  & \ \  $-2.6$  \\
\hline\hline
\end{tabular}
\label{ZP table}
\end{table*}
\begin{table*}[t]
\begin{center} \caption{ \sf Values of input parameters.}
\begin{tabular}{|lr|lr|}
\hline
$ \alpha_{em}(M_Z)                     =1/128.940 $ & \cite{Misiak:2006zs}&
$ \alpha_s(M_Z)                        = 0.1184 \pm 0.0007  $ & \cite{pdg} \\
\hline
$ m_e     = 0.51099 \times 10^{-3}   \ {\rm GeV} $   & \cite{pdg}  &$m_\mu   = 0.10565837   \ {\rm GeV} $ &  \cite{pdg}\\
$m_B=5.27950   \ {\rm GeV}$ & \cite{pdg} & $m_{K^*}=0.89594   \ {\rm GeV}$  & \cite{pdg} \\
$ m_b^{1S} = 4.68 \pm 0.03   \ {\rm GeV}   $  &\cite{Bauer:2004ve} &
$ m_s = 0.095 \pm 0.005 \ {\rm GeV}$ &   \cite{pdg} \\
$ m_c^{\overline{MS}}(m_c)  = 1.27 \pm 0.09  \ {\rm GeV}    $ & \cite{pdg} & & \\
\hline
$ |V_{tb}| = 0.999139\pm 0.000045$ &  \cite{pdg}  & $ |V_{ts}| = (40.5\pm 0.11)\cdot 10^{-3}$ & \cite{pdg}\\
\hline
$ f_B = 194 \pm 10\ {\rm MeV}    $ & \cite{Mahmoudi:2012un} & $\lambda_B = 460 \pm 110 \ {\rm MeV} $& \cite{Ball:2006nr} \\
$f_{K^*,||}= 220 \pm 5$\ {\rm MeV} & \cite{Ball:2006eu} & $f_{K^*,\perp} = 185 \pm 9\ {\rm MeV}$
 \hspace{0.15cm}
 & \cite{Ball:2006eu}\\
$ a_{1,||} = 0.03\pm 0.03$ & \cite{zwicky05}& $ a_{2,||} = 0.08\pm 0.06$ & \cite{zwicky05}\\
$ a_{1,\perp} = 0.03\pm 0.03$ & \cite{zwicky05}& $ a_{2,\perp} = 0.08\pm 0.06$ & \cite{zwicky05}\\
\hline \hline
\end{tabular}
\label{TabInputs1}
\end{center}
\end{table*}

\section{} \label{TransAmp}
The expression of  $\Delta_\parallel$, appear in the definition of a$_{0}$ below Eq. (\ref{EKdef}), written as follows
\begin{eqnarray}
\Delta_\parallel (s) &=& 1 + \frac{ \alpha_s C_F}{4\pi} \left[(2L-2)\right.\notag \\
&&\left.-\frac{2s}{E^2_{K^*}}
\frac{\pi^2 f_B f_{K^*\parallel} \lambda^{-1}_{B +}}{N_c m_B (E_{K^*}/m_{K^*}) \xi_{\parallel} (s)} \;
\int^1_0 \frac{du}{\bar{u}} \Phi_{\bar{K}^*,\parallel}\right],\notag \\\label{delta}
\end{eqnarray}
and contributes only for massive leptons. The light-cone distribution
amplitude (LCDA) $\Phi_{\bar{K}^*,a}$ for transversely
$\left(a = \perp\right)$ and longitudinally $\left(a =\parallel\right)$ polarized $K^*$ can be written as \cite{Beneke:2001at,ball03}
\begin{eqnarray}
 \Phi_{\bar{K}^*,a} &=& 6u\left(1-u\right)\lbrace 1 + a_1 \left(\bar{K}^*\right)_a C_1^{\left(3/2\right)}\left(2u-1\right)\notag \\
&&\qquad+ a_2 \left(\bar{K}^*\right)_a C_2^{\left(3/2\right)}\left(2u-1\right)\rbrace\;,
\end{eqnarray}
where $L = -(m^2_b - s)/s \ln\left( 1-s/m^2_b \right)$ and $a_i \left(\bar{K}^*\right)_a$ are the Gegenbauer coefficients.
The moments are
\begin{eqnarray}
\lambda^{-1}_{B,+} &=& \int_{0}^{\infty} d\omega\frac{\Phi_{B,+}\left(\omega\right)}{\omega},\notag \\
\lambda^{-1}_{B,-} &=& \int_{0}^{\infty} d\omega\frac{\Phi_{B,-}\left(\omega \right)}{\omega- s/m_B -i \epsilon}\; \nn
\end{eqnarray}
where $\Phi_{B,\pm}$ are the two $B$-meson light-cone distribution amplitudes
\cite{Beneke:2001at}. The $\lambda^{-1}_{B,-}\left(s\right)$ can be expressed as:
\begin{equation}
 \lambda^{-1}_{B,-}\left(s\right) = \frac{e^{-s/\left(m_B\omega_0\right)}}{\omega_0}\left[-\textbf{E}\textbf{i}\left(s/m_B\omega_0\right) + i\pi\right], \nn
\end{equation}
where $\omega_0 = 2(m_B - m_b)$. The $\xi_a$ are the universal form factors,
\begin{eqnarray}
\xi_\perp &=& \frac{m_B}{m_B + m_{K^*}} V\notag \\
\xi_\parallel &=& \frac{m_B + m_{K^*}}{2 E_{K^*}} A1 - \frac{m_B - m_{K^*}}{m_B} A2.
\end{eqnarray}

%
%
The $B\rightarrow K^*$ matrix elements in heavy quark limit depend on four independent functions
$\mathcal{T}_a^\pm$ $\left(a = \perp, \parallel\right)$. In the low $s$, ($1.0 < s < 6.0\,$GeV$^2$), the invariant
amplitudes $\mathcal{T}_{\perp,\parallel}$ at NLO within QCDf are given in \cite{hiller2, ball, Beneke:2001at},
\begin{eqnarray}
\mathcal{T}_a &=& \xi_a C_a + \frac{\pi^2}{N_c}\frac{f_B f_{K^* ,a}}{m_B} \Xi_a \sum\limits_\pm \int \frac{d\omega}{\omega}\notag \\
&&\qquad\times
\Phi_{B,\pm}\left(\omega\right) \int_{0}^{1} du \Phi_{K^*,a}\left(u\right) T_{a,\pm}\left(u,\omega\right)\;,\label{B5}
\end{eqnarray}
where $\Xi_\perp \equiv 1$, $\Xi_\parallel \equiv m_{K^*}/E_{K^*}$ and the factorization scale $\mu_f = \sqrt{m_b \Lambda_{QCD}}$.
The coefficient functions $C_a$ and hard scattering functions $T_{a , \pm}$ are written as
\begin{eqnarray}
C_a &=& C_a^{ (0)} +  \frac{\alpha_s\left(\mu_b\right) C_F}{4\pi}C_a^{ (1)}\notag \\
T_{a , \pm} &=& T_{a , \pm}^{(0)} \left(u ,\omega\right) + \frac{\alpha_s\left(\mu_f\right) C_F}{4\pi}
T_{a , \pm}^{(1)} \left(u ,\omega\right).
\end{eqnarray}
 The form factor terms $C_a^{\left(0\right)}$ at LO are
 \begin{equation}
 C_\perp^{ \left(0\right)} = C_7^{\rm eff} + \frac{s}{2m_b m_B} Y\left(s\right)\;,~~~~~{\rm and}~~~~~
 C_\parallel^{ \left(0\right)} = -C_7^{\rm eff} - \frac{ m_B}{2m_b} Y\left(s\right).\nn
 \end{equation}
\begin{align}
Y(s) &= h(s, m_c) \left( \frac{4}{3} C_1 + C_2 + 6 C_3 + 60 C_5\right)\notag \\
&-\frac{1}{2}h(s, m^{\rm pole}_b) \left( 7C_3 +  \frac{4}{3} C_4 + 76 C_5 + \frac{64}{3} C_6\right) \nn \\
&-\frac{1}{2}h(s, 0) \left( C_3 +  \frac{4}{3} C_4 + 16 C_5 + \frac{64}{3} C_6\right)\notag \\
&+\frac{4}{3} C_4 + \frac{64}{9} C_5 + \frac{64}{27} C_6, \nn
\end{align}
where $h(s, m_q)$ is well-known  fermionic loop function.\\
The coefficients $C_a^{\left(1\right)}$ at NLO is divided into a factorizable and a non-factorizable part as
\begin{equation}
C_a^{\left(1\right)} = C_a^{ \left(f\right)} + C_a^{ \left(nf\right)}\;.
\end{equation}
 At NLO the factorizable correction reads \cite{Beneke:2001at, beneke05}
 \begin{eqnarray}
 C_\perp^{ \left(f\right)} &=& C_7^{\rm eff} \left(\ln \frac{m_b^2}{\mu^2}- L + \Delta M \right)\notag \\
 C_\parallel^{\left(f\right)} &=& -C_7^{\rm eff} \left(\ln \frac{m_b^2}{\mu^2}+ 2L + \Delta M \right).\nn
\end{eqnarray}
The non-factorizable corrections are,
\footnotesize
 \beqa
 &&C_F  C_\perp^{ \left(nf\right)} =  -\bar{C}_2 F_2^{\left(7\right)} - C_8^{\rm eff}F_8^{\left(7\right)}\notag \\
 && -\frac{s}{2m_b m_B}
\left[\bar{C_2} F_2^{\left(9\right)} + 2\bar{C}_1\left( F_1^{\left(9\right)}+\frac{1}{6}F_2^{\left(9\right)}\right)
+ C_8^{\rm eff}F_8^{\left(9\right)}\right],\nn\\
 && C_F  C_\parallel^{ \left(nf\right)}=  \bar{C}_2 F_2^{\left(7\right)} + C_8^{\rm eff}F_8^{\left(7\right)}\notag \\
&&+\frac{ m_B}{2m_b} \left[\bar{C_2} F_2^{\left(9\right)} + 2\bar{C}_1\left( F_1^{\left(9\right)}+\frac{1}{6}F_2^{\left(9\right)}\right)
+ C_8^{\rm eff}F_8^{\left(9\right)}\right], \nn
 \eeqa
 \normalsize
where $\Delta M$ depends on the mass renormalization convention for $m_b$.  These corrections are obtained from the matrix
elements of four-quark and chromomagnetic dipole operators \cite{Beneke:2001at}
that are embedded in $F^{(7,9)}_{1,2}$ and $F^{(7,9)}_8$ \cite{asatryan01, greub08}. \\
At LO the hard-spectator scattering term $T_{a , \pm}^{(0)} \left(u ,\omega\right)$ from weak annihilation diagram is \cite{Beneke:2001at}
\begin{equation}
T_{\perp , +}^{(0) } \left(u ,\omega\right) = T_{\perp , -}^{(0) } \left(u ,\omega\right) = T_{\parallel , +}^{(0) }
\left(u ,\omega\right) = 0\;,\nn
\end{equation}
\begin{equation}
\hspace{1cm} T_{\parallel , -}^{(0) } \left(u ,\omega\right) = -e_q \frac{m_B \omega}
{m_B \omega -s -i\epsilon}\frac{4m_B}{m_b} \left(\bar{C}_3 + 3\bar{C_4}\right)\;. \nn
\end{equation}
The contributions to $T_a^{(1)}$ at NLO also contain a factorizable as well as non-factorizable part
\begin{equation}
T_a^{(1)} = T_a^{(f)} + T_a^{(nf)}\;.
\end{equation}
Including $\mathcal{O}\left(\alpha_s\right)$ corrections the factorizable term to $T_{a ,\pm}^{(1)}$
are given by \cite{Beneke:2001at, beneke05}
\begin{eqnarray}
T_{\perp ,+}^{(f) } \left(u, \omega\right) &=& C_7^{\rm eff}\frac{2m_B}{\bar{u}E_{K^*}}\;,~~~
T_{\parallel ,+}^{(f) } \left(u, \omega\right) = C_7^{\rm eff}\frac{4m_B}{\bar{u}E_{K^*}}\notag \\
T_{\perp ,-}^{(f) } \left(u, \omega\right) &=& T_{\parallel ,-}^{(f) } \left(u, \omega\right) = 0\;,\nn
\end{eqnarray}
where $\bar{u} = 1-u$. The non-factorizable correction comes through the matrix elements of
four-quark operators and the chromomagnetic dipole operator
\begin{align}
&T_{\perp,+}^{(nf) } \left(u, \omega\right)  = -\frac{4e_d C_8^{\rm eff}}{u + \bar{u}s/m_B^2} \notag \\
&+ \frac{m_B}{2m_b}[ e_u t_\perp
\left(u, m_c\right) \left(\bar{C_2} + \bar{C_4} - \bar{C_6}\right)\nn \\
& + e_d t_\perp \left(u, m_b\right) \left(\bar{C_3} + \bar{C_4} - \bar{C_6}-4m_b/m_B \bar{C_5}\right)\notag \\
& +
e_d t_\perp \left(u, 0\right)\bar{C_3} ], \nn\\
&T_{\perp,-}^{(nf) } \left(u, \omega\right) = 0\;,\nn \\
&T_{\parallel,+}^{(nf) } \left(u, \omega\right)  =  \frac{m_B}{m_b} [ e_u t_\parallel \left(u, m_c\right) \left(\bar{C_2}
 + \bar{C_4} - \bar{C_6}\right) \notag \\
 &+ e_d t_\parallel \left(u, m_b\right) \left(\bar{C_3} + \bar{C_4} - \bar{C_6} \right)
+ e_d t_\parallel \left(u, 0\right)\bar{C_3} ]\;,\nn \\
&T_{\parallel,-}^{(nf) } \left(u, \omega\right)  = e_q \frac{m_B\omega}{m_B\omega - s - i\epsilon} \Bigg[
\frac{8C_8^{\rm eff}}{\bar{u} + us/m_B^2} \notag \\
&+ \frac{6m_B}{m_b} \Bigg(h\left(\bar{u}m_B^2 + us , m_c\right)\left(\bar{C}_2 +
\bar{C}_4 + \bar{C}_6\right)\nn \\
 &+ h\left(\bar{u}m_B^2 + us , m_b^{pole}\right)\left(\bar{C}_3 + \bar{C}_4 + \bar{C}_6\right) \notag \\
 &+ h\left(\bar{u}m_B^2 + us , 0\right)\left(\bar{C}_3 + 3\bar{C}_4 + 3\bar{C}_6\right)\nn \\
&-\frac{8}{27} \left(\bar{C}_3 - \bar{C}_5 - 15\bar{C}_6\right)\Bigg) \Bigg]\nn.
\end{align}
The $t_a \left(u ,m_q\right)$ functions are given by
\beqa
t_\perp \left(u ,m_q\right) &=& \frac{2m_B}{\bar{u}E_{K^*}}I_1\left(m_q\right) + \frac{s}{\bar{u}^2E^2_{K^*}}\notag \\
&&\times\left(B_0\left(\bar{u}m_B^2 + us , m_q \right)
 - B_0\left(s , m_q\right)\right),\nn\\
t_\parallel \left(u ,m_q\right) &=&  \frac{2m_B}{\bar{u}E_{K^*}}I_1\left(m_q\right) + \frac{\bar{u}m_B^2 + us}{\bar{u}^2E^2_{K^*}}\notag \\
&&\times\left(B_0\left(\bar{u}m_B^2 + us , m_q \right)- B_0\left(s , m_q\right)\right)\;,\nn
\eeqa
where $B_0$ and $I_1$ are
\begin{equation}
B_0\left(s ,m_q\right) = -2\sqrt{4m^2_q/s - 1} \arctan \frac{1}{\sqrt{4m^2_q/s - 1}}\;,\nn\hspace{2.3cm}
\end{equation}
\footnotesize
\begin{equation}
I_1\left(m_q\right) = 1 + \frac{2m^2_q}{\bar{u}\left(m^2_B - s\right)}\left[L_1\left(x_+\right) + L_1\left(x_-\right) - L_1\left(y_+\right)
- L_1\left(y_-\right)\right]\;,\nn
\end{equation}
\normalsize
and
\footnotesize
\begin{equation}
x_\pm = \frac{1}{2} \pm \left(\frac{1}{4} - \frac{m^2_q}{\bar{u}m_B^2 + us}\right) ^{1/2}, y_\pm
 = \frac{1}{2} \pm \left(\frac{1}{4} - \frac{m^2_q}{s}\right)^{1/2}\;, \nn
\end{equation}
\normalsize
\begin{equation}
L_1\left(x\right) = \ln \frac{x-1}{x} \ln\left(1-x\right) - \frac{\pi^2}{6} + Li_2\left(\frac{x}{x-1}\right).\nn \hspace{2.3cm}
\end{equation}

%

\end{document}